\documentclass[AMA,Times1COL]{WileyNJDv5}

\articletype{Research Article}%

\received{Date Month Year}
\revised{Date Month Year}
\accepted{Date Month Year}
\journal{Journal}
\volume{00}
\copyyear{2026}
\startpage{1}

\usepackage{optidef}
\usepackage{amsmath}
\DeclareMathOperator*{\argmax}{arg\,max}
\usepackage{subcaption}
\usepackage{graphicx}

\usepackage{xcolor}

\makeatletter
\def\oddhead@titlepage@info{}\def\evenhead@titlepage@info{}
\def\oddfoot@titlepage@info{}\def\evenfoot@titlepage@info{}
\makeatother
\jnlcitation{}

\begin{document}

\title{Gain-function optimisation of graphical multiple testing procedures for confirmatory clinical trials}

\author[1]{Alexander D.V. Spiers}

\author[2]{Michael J. Grayling}

\author[1]{Graham M. Wheeler}

\author[1]{Adrian P. Mander}

\authormark{SPIERS \textsc{et al.}}
\titlemark{Gain-function optimisation of graphical multiple testing procedures for confirmatory clinical trials}

\address[1]{\orgdiv{Quantitative Sciences Innovation}, \orgname{GSK}, \orgaddress{\state{London}, \country{UK}}}

\address[2]{\orgdiv{Statistics and Decision Sciences}, \orgname{Johnson \& Johnson}, \orgaddress{\state{High Wycombe}, \country{UK}}}

\corres{Corresponding author Alexander Spiers. \email{alex.d.spiers@gsk.com}}


\abstract[Abstract]{Graphical multiple testing procedures are a flexible and transparent way to control the family-wise error rate when a confirmatory trial pursues several label claims, but they leave open the question of which graph to use.
In practice sponsors often fall back on fixed-sequence or Holm procedures that may poorly reflect what the trial is actually trying to achieve.
We propose to instead choose the graph that maximises an explicit gain function, which states what each possible set of rejected hypotheses is worth to the sponsor: typically nothing until a regulatory hurdle is cleared, then an incremental amount for each further claim.
Conventional power criteria are recovered as special cases, and because the search is confined to graphical procedures, family-wise error rate control holds whichever graph is selected.
We set out a structured procedure for eliciting the gain function from the clinical, commercial, and regulatory members of a trial team.
The proposed framework also handles uncertainty in the treatment effects and correlations assumed at the design stage, by averaging performance over their plausible values rather than fixing a single assumption.
For group sequential designs, we describe how it can further reward early claims.
Five examples, drawn from real and hypothetical pharmaceutical trials, show that the best graph depends on how trial success is defined, how uncertain the design assumptions are, and, in a group sequential setting, when claims can be established.}


\jnlcitation{\cname{%
\author{Spiers A},
\author{Grayling M},
\author{Wheeler G}, and
\author{Mander AP}}.
\ctitle{Gain-function optimisation of graphical multiple testing procedures for confirmatory clinical trials.} \cjournal{\it Journal Name.} \cvol{2021;00(00):1--18}.}

\maketitle

\renewcommand\thefootnote{}
\footnotetext{\textbf{Abbreviations:} BMD, bone mineral density; DLQI, Dermatology Life Quality Index; DR, direct rating; EASI, Eczema Area and Severity Index; FVC, forced vital capacity; FWER, family-wise error rate; GSD, group sequential design; IGA, Investigator's Global Assessment; OS, overall survival; PFS, progression-free survival; SCORAD, SCORing Atopic Dermatitis}

\renewcommand\thefootnote{\fnsymbol{footnote}}
\setcounter{footnote}{1}

\section{Introduction} \label{sec:intro}

Clinical trials in drug development often include multiple objectives to comprehensively assess the efficacy and safety of new treatments.
Testing these objectives together gives rise to a multiple testing problem that typically necessitates a multiple comparisons procedure.
Multiple comparisons can arise from various sources, including the assessment of numerous primary and secondary endpoints, comparisons of several treatments or doses against a control, subgroup analyses, or repeated assessment of endpoints over time.\cite{dmitrienko2013, dmitrienko2018}
If each individual hypothesis test is conducted at its nominal level without adjustment, the probability of making at least one false rejection increases with the number of null hypotheses tested.
In confirmatory trials, when multiple claims are intended to support product labelling, strong control of the family-wise error rate (FWER) is expected by regulatory agencies as the standard for limiting spurious claims.\cite{EMA2016, FDA2022}
Within this framework, hypotheses are commonly organised into primary and key secondary families:
demonstrating efficacy on primary endpoint(s) is typically required for regulatory approval, while key secondary endpoints may support additional label claims that enhance the clinical and commercial value of the product.\cite{dmitrienko2018, EMA2016, FDA2022}

Graphical approaches to multiple testing procedures, introduced by Bretz et al. \cite{bretz2009} and Burman et al.,\cite{burman2009} provide a visually transparent framework for controlling the FWER.
Such graphs are simpler to communicate than more general closed testing procedures, while retaining most commonly used multiplicity strategies as special cases.
For example, this class includes fixed-sequence (hierarchical),\cite{wiens2003} Bonferroni-Holm,\cite{holm1979} fallback,\cite{wiens2005} and gatekeeping procedures \cite{dmitrienko2003, dmitrienko2008, xi2014} as special cases, all of which adhere to the closure principle and thus control the FWER in the strong sense.\cite{marcus1976}
In some settings the testing order will be dictated by dose strength or the timing of endpoints; a fixed-sequence procedure is then a natural choice. When such a rationale is absent, many sponsors still default to fixed-sequence procedures because they are simple to communicate and implement.\cite{wiens2003}
However, with this approach, a misspecified testing order can cause the procedure to stop before reaching a hypothesis with a small nominal p-value, forfeiting a valuable label claim.\cite{dmitrienko2018}
This creates pressure to order hypotheses by their likelihood of achieving significance rather than by the value of the resulting claims, potentially misaligning the testing strategy with the trial's objectives.
General graphical procedures address this limitation by allowing multiple testing pathways through \textit{alpha-recycling}, so that a single failed hypothesis need not block the entire procedure.
However, the question of how to specify the graph so that the procedure is transparently aligned with the value of different label claims remains open.

Several authors have proposed optimising graphical procedures with respect to power-based objectives.\cite{millen2011, zhang2015, zhan2022, zhang2023, xi2024}
In particular, Xi and Chen \cite{xi2024} optimised the hypothesis weights of weighted Bonferroni tests to maximise disjunctive and conjunctive power.
They extended their work to graphical approaches via a two-step approach: the initial weights were first optimised for disjunctive power, then the transition matrix was determined by re-optimising the weights after each hypothetical single rejection.
Zhan et al. \cite{zhan2022} used deep learning to maximise a weighted sum of multiplicity-adjusted powers over the graph parameters, where user-specified weights reflected the relative importance of each hypothesis.
These approaches provide important benchmarks, but evaluate procedures through a single power metric: disjunctive, conjunctive, or weighted-average power.
In practice, a trial delivers no regulatory or commercial value unless the evidence required for regulatory submission is achieved (e.g., rejection of all co-primary endpoints);
once this hurdle is cleared, each additional rejection may support a further claim in the product label, increasing the drug's commercial value.
Conventional power metrics do not capture this conditional structure.
Furthermore, existing methods optimise for a single fixed alternative and do not account for uncertainty in treatment effects and correlations at the design stage.
Both of these limitations have been identified previously as important open problems.\cite{zhang2023, xi2024}

Lisovskaja and Burman \cite{lisovskaja2015} proposed a decision-theoretic alternative approach. 
They defined a utility function of a vector of rejection indicators and maximised expected utility over the full class of Bonferroni-based closed testing procedures.
Their results showed that optimal procedures under general utility functions can differ substantially from those based on standard power metrics, and depend heavily on the assumed effect sizes and correlations.
However, their exact computation of expected utility is tractable only under independent test statistics, and they did not address how to elicit their utility functions in practice.
A further issue is that effect sizes and correlations are uncertain at the design stage; therefore one might prefer to evaluate procedures across a range of plausible scenarios rather than condition on a single fixed alternative.
Within a broader framework for controlling family-wise expected loss, Maurer et al. \cite{maurer2023} formalised the \textit{Bayes gain}, defined as the expected gain averaged over a prior on the test statistic distribution, and established its equivalence to optimising under a single predictive distribution for the p-values.
However, they did not demonstrate how it may be applied to graphical procedure optimisation in practice.

Accordingly, this paper develops these decision-theoretic ideas into a practical framework for optimising graphical multiple testing procedures in confirmatory clinical trials.
First, we introduce a family of gain functions that assign zero value unless a minimum regulatory requirement is met, and award incremental value for each additional rejection beyond that threshold -- standard power criteria arise as special cases.
We use the expected gain as an objective function to optimise graphical procedures directly, replacing the power-based objectives of previous approaches with an objective that reflects the conditional structure of trial success.
Second, we propose a structured elicitation procedure for the gain function parameters, enabling cross-functional trial-planning teams to translate clinical, commercial, and regulatory priorities into the objective function.
Third, we use the Bayes gain as an alternative objective function for graphical procedure optimisation and extend the framework to group sequential designs (GSDs) through time-dependent gain functions,
yielding procedures that account for both design-stage uncertainty and the timing of confirmatory claims.

Section~\ref{sec:methods} formalises the expected gain of a graphical procedure, introduces the hurdle-structured gain function and its elicitation, defines the Bayes gain, and presents the optimisation formulations for fixed-sample and GSDs. 
Section~\ref{sec:examples} illustrates the framework through five examples based on real and hypothetical pharmaceutical trials. Section~\ref{sec:discussion} discusses implications for practice and directions for future work.

\section{Methods}\label{sec:methods}

\subsection{Distributional assumptions}\label{sec:distrib_assumptions}

Consider a clinical trial that tests a family $\mathcal{H}$ of $m$ one-sided null hypotheses:
\[
H_i : \mu_i \leq 0 \quad\text{versus}\quad H^A_i : \mu_i > 0, \quad i \in \{1,\dots,m\}.
\]
We summarise the data for hypothesis $H_i$ by a test statistic $Z_i$,
and the associated one-sided p-value
\[
p_i=1-\Phi\left(Z_i\right), \quad i=1, \ldots, m,
\]
where $\Phi$ denotes the standard normal distribution function. Let
\[
\boldsymbol{Z}=\left(Z_1, \ldots, Z_m\right)^{\intercal}, \quad \mathbf{p}=\left(p_1, \ldots, p_m\right)^{\intercal} .
\]
Throughout, we will assume that under a fixed global alternative the vector of test statistics 
$\boldsymbol{Z} = (Z_1,\dots,Z_m)^\intercal$
asymptotically follows a multivariate normal distribution 
$
\boldsymbol{Z} \sim \mathcal{N}_m(\boldsymbol{\Delta}, \boldsymbol{\Sigma})
$, where: (i) $\boldsymbol{\Delta} = (\Delta_1, \dots, \Delta_m)^\intercal$ is the mean vector of the test statistics under the assumed alternatives (noncentrality parameters); and (ii) $\boldsymbol{\Sigma}$ is the $m \times m$ variance-covariance matrix of the test statistics, which also serves as the correlation matrix, since $Z_i$ is constructed to have an asymptotic variance of 1. We collect all parameters governing the joint test statistic distribution with $\boldsymbol \theta = (\boldsymbol{\Delta}, \boldsymbol{\Sigma}) \in \boldsymbol \Theta$.
The corresponding joint density of the p-values is denoted by $f(\mathbf{p} \mid \boldsymbol \theta)$.

\subsection{Graphical multiple testing procedures}\label{sec:graphical_testing}

Bonferroni-based graphical procedures use weighted Bonferroni intersection tests and provide a sequentially rejective shortcut to the corresponding closed testing procedure; they thus strongly control the FWER at level $\alpha$.\cite{bretz2009, burman2009}
As with more general closed testing procedures, graphical procedures can be tailored with weights to quantify the relative importance of the hypotheses, but are often easier than the former to visualise and communicate to clinical teams.
Each graphical procedure is uniquely determined by two components. 
The first is the initial hypothesis weight vector $\mathbf{w} = (w_1,\dots,w_m)$ with $\sum_{i = 1}^m w_i \leq 1$, which defines the \textit{alpha-splitting}: each hypothesis $H_i$ receives an initial local significance level $w_i \alpha$.
The second is a transition matrix $\mathbf{G} = (g_{ij})_{i,j=1,\ldots,m}$, which defines the \textit{alpha-recycling}: upon rejection of $H_i$, the fraction $g_{ij}$ of its local significance level is redistributed to $H_j$.
The matrix $\mathbf{G}$ satisfies constraints $0 \leq g_{ij} \leq 1$, $\sum_{j=1}^m g_{ij} \leq 1$, and $g_{ii}=0$ for all $i$.

Graphical procedures using weighted-Bonferroni local tests follow a sequentially rejective algorithm. At each step, any hypothesis $H_j$ whose p-value satisfies $p_j \le w_j(I)\alpha$ is rejected, where $w_j(I)$ is the current weight for $H_j$ given the set $I \subseteq \{1, 2, \dots, m\}$ of hypotheses not yet rejected.
The local significance level is then alpha-recycled to the remaining hypotheses according to $\mathbf{G}$. The transition weights and $I$ are then updated.
The procedure terminates when no further rejections are possible.
Full details, including the update equations for $w_j(I)$ and $g_{ij}$, are given in Bretz et al. \cite{bretz2009} (see Algorithm 1).

All optimisation in this paper is conducted over Bonferroni-based graphical procedures. As such, strong FWER control is a guaranteed property of the procedure itself and holds regardless of the joint distribution of the test statistics,
even if the working multivariate model is misspecified.

\paragraph*{Extension of graphical approaches to group sequential designs}
\vspace*{6pt}

Maurer and Bretz \cite{maurer2013} extended this framework to GSDs, where hypotheses are tested repeatedly at $K$ pre-planned analyses.\cite{jennison1999}
Because endpoints may accrue information at different rates, let $t_{i,k}\in[0,1]$ denote the information fraction of hypothesis $H_i$ at analysis $k$, where
$
0\le t_{i,1}\le t_{i,2}\le\cdots\le t_{i,K}=1.
$
The information fractions are hypothesis-specific and non-decreasing in $k$.
Each hypothesis $H_i$ is assigned a non-decreasing error-spending function $\phi_i(a,t)$ with $\phi_i(a,0)=0$ and $\phi_i(a,1)=a$, giving the cumulative amount of the allocated local level $a$ that may be spent by information fraction $t$. 
The per-analysis increments of $\phi_i$, together with the joint distribution of the sequential test statistics, determine the nominal rejection boundaries $\alpha^*_{i,k}(a)$.

The group sequential graphical procedure operates on the initial graph $(\mathbf{w}, \mathbf{G})$ using the sequential rejection and alpha-recycling rules of the fixed-sample procedure, adapted to accommodate repeated analyses.
As in the fixed-sample case, let $I$ index the hypotheses not yet rejected and $w_j(I)$ the current weight for $H_j$.
At analysis $k$, the fixed-sample rejection condition $p_j \leq w_j(I)\alpha$ is replaced by $p_{j,k} \leq \alpha^*_{j,k}(w_j(I)\alpha)$, where $\alpha^*_{j,k}(a)$ is the nominal boundary determined by the spending function $\phi_j$ at information fraction $t_{j,k}$ and local significance level $a$.
When a hypothesis is rejected, the graph is updated by alpha-recycling according to $\mathbf{G}$, and the new weights define the spending function boundaries both at the current analysis and at all subsequent analyses.
This ensures strong FWER control across all hypotheses and analyses.

\subsection{Expected gain as an objective function}

\subsubsection{Defining gain functions for graphical procedures} \label{sec:define_gain_function}

Given observed p-values, $\mathbf{p}$, denote by $R$ the graphical procedure that determines which hypotheses are rejected according to the selected $(\mathbf{w}, \mathbf{G})$.
We write the outcome as
\[
\mathbf{r} = R(\mathbf{p};\, \mathbf{w}, \mathbf{G}) \in \{0,1\}^m,
\]
where $\mathbf{r} = (r_1, ..., r_m)$ denotes what we term the \textit{rejection pattern}: $r_i = 1$ if $H_i$ is rejected and $r_i = 0$ otherwise.
We assume that the overall target FWER $\alpha$ is fixed henceforth. 
Under the assumed joint distribution described in Section~\ref{sec:distrib_assumptions}, both $\mathbf{p}$ and $\mathbf{r}$ are random, and their distribution 
depends on the test statistic parameters 
$\boldsymbol \theta = (\boldsymbol{\Delta},\boldsymbol{\Sigma})$
and the choice of the graph parameters $(\mathbf{w},\mathbf{G})$.
Our aim is to choose parameters $(\mathbf{w},\mathbf{G})$ --- and hence the 
rejection rule $R$ ---
so that the resulting distribution of $\mathbf{r}$ is as favourable as possible with respect to the objectives of the planned trial, which we encode in a prespecified gain function. Following Lisovskaja and Burman \cite{lisovskaja2015} who optimised a utility function over rejection patterns, we define a gain function 
\[
\psi : \mathcal{R} \to \mathbb{R}_{\geq 0},
\]
where $\mathcal{R} = \{0,1\}^m$ is the set of all possible rejection patterns.
We use the term \textit{gain} rather than \textit{utility} to avoid confusion with health-economic utilities.
The framework places no restriction on $\psi$: any non-negative function of the rejection pattern is permissible.
Different choices of $\psi$ simply encode different trial priorities.
Conventional power metrics are one family of special cases:

\begin{itemize}
    \item \textbf{Disjunctive Power} (rejecting at least one hypothesis / ``\textit{multiple shots on goal}''):
    $$
        \psi(\mathbf{r}) = 
        \begin{cases}
        1 & \text{if } \sum_{i=1}^{m} r_i \geq 1, \\
        0 & \text{otherwise};
        \end{cases}
    $$
    \item \textbf{Conjunctive Power} (rejecting all hypotheses simultaneously / ``\textit{all or nothing}''):
    $$
        \psi(\mathbf{r}) = 
        \begin{cases}
        1 & \text{if } \sum_{i=1}^{m} r_i = m, \\
        0 & \text{otherwise};
        \end{cases}
    $$

    \item \textbf{Average Power} (expected proportion of rejected hypotheses):

    $$
        \psi(\mathbf{r}) = \frac{1}{m}\sum_{i=1}^{m} r_i.
    $$
\end{itemize}
These metrics treat all rejections symmetrically and do not distinguish between regulatory requirements and supporting label claims.
In confirmatory trials, however, value typically accrues conditionally: none is realised unless a minimum set of requirements is met (such as rejection of a primary endpoint), and each additional rejection adds value only thereafter.
Therefore one useful family of gain functions has this \textit{hurdle-structure}:
\begin{equation}\label{eq:hurdle_formula}
\psi(\mathbf{r})=\mathbb{I}_{\text {reg}}(\mathbf{r}) \cdot\left(v_{\text {base}}+\psi_{\text {incr}}(\mathbf{r})\right).
\end{equation}
\begin{itemize}
    \item $\mathbb{I}_{\text {reg}}(\mathbf{r})$ is the regulatory success indicator, taking the value 1 if the rejection configuration $\mathbf{r}$ meets the minimum criteria for drug approval, and 0 otherwise. $\mathbb{I}_{\text {reg}}(\mathbf{r})$ could encode any Boolean success criterion. For example, approval may require both co-primary endpoints to be met ($\mathbb{I}_{\text{reg}}(\mathbf{r})=r_1r_2$), at least one of two dual-primary endpoints ($\mathbb{I}_{\text{reg}}(\mathbf{r})=\mathbb{I}\{r_1+r_2\ge1\}$), or a more elaborate gatekeeping rule. The choice of function will be determined by regulatory expectations on dual vs co-primary endpoints. \cite{liu2026}
    \item $v_{\text {base}}$ represents the base value of achieving regulatory approval. In practice, $v_{\text{base}}$ might represent the projected annual revenue from regulatory approval (e.g., \$200M).
    \item $\psi_{\text {incr}}(\mathbf{r})$ is the incremental gain function, representing the incremental value of additional label claims, which contributes to the total gain only if $\mathbb{I}_{\text {reg}}(\mathbf{r})=1$.
\end{itemize}
This structure is also convenient for the elicitation procedures we propose later (see Section \ref{sec:methods_elicitation}).
Throughout this paper we normalise so that $\psi(\mathbf{r}) \in [0,1]$, but the framework applies equally to absolute monetary or market-share scales. Because the base and incremental values are summed, they must be measured on a common scale; the choice of scale is free, since multiplying $\psi$ by a positive constant does not change the optimal graph.

\paragraph*{Example of gain function in an osteoporosis trial}
\vspace*{6pt}

Consider a confirmatory phase~III trial evaluating a novel therapy for postmenopausal osteoporosis against control.
The trial aims to demonstrate efficacy through one primary and two supportive secondary endpoints.
The hypotheses tested are:

\begin{itemize}
    \item $H_1$ (primary): reduction in the incidence of new vertebral fractures over 24 months.
    \item $H_2$ (key secondary): time to first hip fracture.
    \item $H_3$ (key secondary): change from baseline in total hip bone mineral density (BMD).
\end{itemize}
Vertebral fracture reduction ($H_1$) forms the primary basis for regulatory approval, while demonstrated reductions in hip fractures ($H_2$) and improvements in BMD ($H_3$) serve as supportive evidence for labelling. \cite{kehoe2019} 
Rejection of $H_2$ or $H_3$ without rejection of $H_1$ yields no additional value for the trial; that is, $\psi(\mathbf{r}) = 0$ whenever $r_1 = 0$. Mapping this to our hurdle structure \eqref{eq:hurdle_formula} gives the gain function:
\begin{equation}
    \psi(\mathbf{r})=r_1 \cdot\left(v_{\text {base}}+v_2 r_2+v_3 r_3\right).
\end{equation}
For instance, if baseline approval is projected to generate \$100M annually and achieving all key secondary endpoints adds a further \$50M through additional label claims, the gain values might be $v_{\text{base}} = 100$, with $v_2 = v_3 = 25$, splitting the \$50M in proportion to their clinical and commercial value. 
The elicitation procedure in Section~\ref{sec:methods_elicitation} provides a structured method to determine these proportions.

\subsubsection{Conditional expected gain for a given test statistic distribution}\label{sec:expected_gain}

We define the \emph{conditional expected gain} of $(\mathbf{w}, \mathbf{G})$ at 
$\boldsymbol{\theta}$ as
\begin{equation}\label{eq:expected_gain} 
U(\mathbf{w}, \mathbf{G} ; \boldsymbol{\theta})=\int \psi\{R(\mathbf{p} ; \mathbf{w}, \mathbf{G})\} f(\mathbf{p} \mid \boldsymbol{\theta}) d \mathbf{p} 
= \mathbb{E}_{\mathbf{p} | \boldsymbol \theta} \bigl[ \psi \left\{ R(\mathbf{p} ; \mathbf{w}, \mathbf{G}) \right\} \bigr].
\end{equation}
Equivalently, summing over the set of all rejection patterns $\mathcal{R}$,
\[
U(\mathbf{w}, \mathbf{G}; \boldsymbol \theta)
=
\sum_{\mathbf{r} \in \mathcal{R}}
\psi(\mathbf{r}) \;
\Pr_{\boldsymbol \theta}\left\{R(\mathbf{p}; \mathbf{w}, \mathbf{G}) = \mathbf{r}\right\}.
\]
The expected gain provides an objective function for comparing and optimising graphical procedures under a fixed test statistic distribution.

Equation~\eqref{eq:expected_gain} can in principle be computed exactly, since each rejection pattern, $\mathbf{r}$, corresponds to a region of test-statistic space that is a finite union of disjoint boxes with known multivariate normal probabilities under the assumptions from Section~\ref{sec:distrib_assumptions}. \cite{zhang2023}
The number of boxes grows very quickly with $m$, however, so we estimate the expected gain by Monte Carlo simulation throughout.
\cite{zhan2022} 
It is calculated by generating $N_{\text{sim}}$ independent draws of p-value vectors $\mathbf{p}^{(b)}=(p_1^{(b)},\dots,p_m^{(b)})^\intercal$, $b=1,\dots,N_{\text{sim}}$, from the assumed fixed-alternative distribution ---
i.e., $\mathbf{p}^{(b)}\sim f(\cdot\mid\boldsymbol\theta)$, where $p_i^{(b)}=1-\Phi(Z_i^{(b)})$ is the p-value for hypothesis $H_i$ in the $b$th draw ---
and applying the graphical procedure $R(\cdot,;\mathbf{w},\mathbf{G})$ to each.
The conditional expected gain is then estimated by the mean gain across the simulated p-value vectors:
$$
\widehat{U}(\mathbf{w},\mathbf{G};\boldsymbol\theta)=\frac{1}{N_{\text{sim}}}\sum_{b=1}^{N_{\text{sim}}}\psi\left\{R(\mathbf{p}^{(b)};\mathbf{w},\mathbf{G})\right\}.
$$
The random variables $\psi\left\{R(\mathbf{p}^{(b)} ; \mathbf{w}, \mathbf{G})\right\}$ are independent and identically distributed, hence this estimator converges to the true conditional expected gain, $U(\mathbf{w}, \mathbf{G};\boldsymbol \theta)$ as $N_{\text{sim}} \to \infty$.

\subsubsection{\textit{Bayes gain}: Incorporating uncertainty in treatment effects and correlation}\label{sec:bayes_gain}

At the design stage of a trial, the true treatment effects and correlation structure are unknown.
Rather than optimising for a single fixed alternative $\boldsymbol{\theta}$, we build on the \textit{Bayes gain} formulation of Maurer et al.,\cite{maurer2023}
which averages the expected gain over a prior distribution on $\boldsymbol{\theta}$ to account for this uncertainty.
We place a prior $\pi(\boldsymbol\theta)$ on the parameter space $\boldsymbol \Theta$.
For any fixed $\boldsymbol \theta$, the conditional expected gain is $U(\mathbf{w}, \mathbf{G}; \boldsymbol \theta)$. We define the Bayes gain, $U_B(\mathbf{w}, \mathbf{G})$ of a graph $(\mathbf{w}, \mathbf{G})$ as:

\begin{equation}\label{eq:bayes_gain}
    U_B(\mathbf{w}, \mathbf{G})=\int_{\Theta} U(\mathbf{w}, \mathbf{G} ; \boldsymbol{\theta}) \pi(\boldsymbol{\theta}) d \boldsymbol{\theta} = \mathbb{E}_\pi[U(\mathbf{w}, \mathbf{G}; \boldsymbol \theta)],    
\end{equation}
where $\mathbb{E}_\pi$ denotes the expectation with respect to the prior $\pi$.
A convenient and interpretable choice for $\pi$ includes finite mixture priors, where probabilities are assigned to finite set of parameter scenarios.
Alternatively, $\pi$ may be taken as a continuous multivariate prior on $(\boldsymbol{\Delta}, \boldsymbol{\Sigma})$, for example, by combining a multivariate prior for treatment effects with a structured prior for the correlation matrix (see Section~S1.2 in Supporting Information).

Following Maurer et al., \cite{maurer2023} we can equivalently express the Bayes gain as an expectation with respect to a single predictive distribution of the p-values. Define the predictive (compound) density of $\mathbf{p}$ by
\[
f^*(\mathbf{p})=\int_{\Theta} f(\mathbf{p} \mid \boldsymbol{\theta}) \pi(\boldsymbol{\theta}) d \boldsymbol{\theta} .
\]
Substituting the expression for $U(\mathbf{w}, \mathbf{G}; \boldsymbol{\theta})$ from Equation~\eqref{eq:expected_gain} into Equation~\eqref{eq:bayes_gain} and changing the order of integration yields,
\[
U_B(\mathbf{w}, \mathbf{G})=\int \psi\left\{R(\mathbf{p} ; \mathbf{w}, \mathbf{G})\right\} f^*(\mathbf{p}) d \mathbf{p}=\mathbb{E}_{f^*}\bigl[\psi\left\{R(\mathbf{p} ; \mathbf{w}, \mathbf{G})\right\}\bigr].
\]
Thus maximising the Bayes gain is equivalent to maximising the expected gain under a single predictive distribution $f^*$ for the p-values
(Proposition 3 of Maurer et al. \cite{maurer2023}).
This collapses what would otherwise be a nested simulation over $\boldsymbol{\theta}$ into a single set of draws from the predictive distribution, so the Bayes gain can be approximated by:
\begin{enumerate}
\item generating random draws $\boldsymbol{\theta}^{(b)}$ from the prior $\pi(\boldsymbol{\theta})$;
\item for each draw, simulating $\mathbf{p}^{(b)}\sim f(\mathbf{p}\mid\boldsymbol{\theta}^{(b)})$;
\item evaluating $\psi\left\{R(\mathbf{p}^{(b)};\mathbf{w},\mathbf{G})\right\}$;
\item averaging to yield:
$$
\widehat{U}_B(\mathbf{w},\mathbf{G})=\frac{1}{N_{\text{sim}}}\sum_{b=1}^{N_{\text{sim}}}\psi\left\{R(\mathbf{p}^{(b)};\mathbf{w},\mathbf{G})\right\}.
$$
\end{enumerate}
The estimator $\widehat{U}_B$ converges to the true Bayes gain as $N_{\text{sim}} \rightarrow \infty$.
It accounts simultaneously for (i) uncertainty in the true joint test-statistic distribution $\boldsymbol{\theta}$ through the prior $\pi$, and (ii) sampling variation in $\mathbf{p}$ for a given $\boldsymbol{\theta}$.

\subsubsection{Time-dependent gain for group sequential designs} \label{sec:gsd_gain}

In GSDs, the gain of a rejected hypothesis may depend on the time of its rejection.
Rejecting a hypothesis at an interim analysis will be more valuable to a sponsor than at the final analysis because it enables earlier patient access and extends the effective patent life of a new treatment.
Let a trial be monitored at $K$ pre-planned analyses with information fractions $0\le t_{i,1}\le t_{i,2}\le\cdots\le t_{i,K}=1$ conducted at calendar times 
$0 < s_1 < \cdots < s_K$ (e.g., months from first patient randomised). 
The outcome of a group sequential graphical procedure can be summarised by: (i) a binary rejection vector $\mathbf{r} \in \{0,1\}^m$, and (ii) a vector of decision times:
\[
\boldsymbol{\tau} = (\tau_1, \ldots, \tau_m), \quad \tau_i \in \{s_1, \ldots, s_K, \infty\}.
\]
Here, $\tau_i$ is the calendar time of the first analysis at which the procedure declares $H_i$ rejected ($\tau_i = \infty$ if $H_i$ is never rejected).

If a hypothesis $H_i$ has data that are already complete at analysis $k$, then $t_{i,k}=1$; its information fraction, and hence its test statistic, does not change thereafter, so $t_{i,k'}=1$ and $p_{i,k'}=p_{i,k}$ for all $k'\ge k$.
This occurs, for example, when a surrogate endpoint matures before the final analysis.
Even if $H_i$ is not rejected at analysis $k$, rejection may occur at a later analysis $k' > k$ if alpha-recycling from other rejections increases $w_i(I)$.
When this happens, the spending function is re-evaluated at the updated local significance level $w_i(I)\alpha$. The resulting boundaries are recomputed at every analysis up to $k'$ (including analysis $k$, where the data originally matured).
This corresponds to the \textit{sequential p-value} framework in which rejection decisions at analysis $k'$ incorporate evidence from all preceding analyses.\cite{zhao2025}
The recorded decision time is $\tau_i = s_{k'}$: it reflects when the rejection criterion was met through alpha-recycling, not when the data matured (see Section~\ref{sec:gsd_example} for an example).

We define the time-dependent gain function:
\[
\psi:\{0,1\}^m \times\{s_1, \ldots, s_K, \infty\}^m \rightarrow \mathbb{R}_{\geq 0},
\]
assigning value to each realised pair $(\mathbf{r}, \boldsymbol{\tau})$. Consistent with the structure introduced in Section~\ref{sec:define_gain_function}, a natural approach is to decompose this gain into a regulatory hurdle and a conditional value component:

\[
\psi(\mathbf{r}, \boldsymbol{\tau})=\mathbb{I}_{\mathrm{reg}}(\mathbf{r}) \cdot \left\{ v_{\text {base}} \cdot d(\tau_{\mathrm{reg}})+\psi_{\text {incr}}(\mathbf{r}, \boldsymbol{\tau})\right\}
\]
where $d: \mathbb{R}_{\geq 0} \cup \{\infty\} \to [0,1]$ is a non-increasing discount function of calendar time, with $d(\infty) = 0$. 
The calendar-time parameterisation ensures that the discount reflects the consequences of delayed regulatory decisions: i.e., deferred patient access, reduced effective patent life, and delayed revenue.
For example, suppose a competitor is expected to complete its pivotal readout at calendar time $s^* = 36$ months.
If first-to-market status captures the full commercial value and second-to-market captures only a fraction $\delta$ (e.g., $\delta = 0.5$),
a simple step function captures this directly:
$d(\tau) = 1$ if $\tau \leq s^*$ and $d(\tau) = \delta$ if $\tau > s^*$.

For fixed information times and spending functions, and under the test statistic distribution scenario $\boldsymbol{\theta}$, we evaluate a candidate graph ($\mathbf{w}, \mathbf{G}$) by its conditional expected gain:
\[
U(\mathbf{w}, \mathbf{G} ; \boldsymbol{\theta})=\mathbb{E}_{\mathbf{p} \mid \boldsymbol{\theta}}[\psi(\mathbf{r}, \boldsymbol{\tau})].
\]
This expectation is also evaluated using Monte Carlo simulation. For each simulation replicate, we generate the joint group sequential test statistics across hypotheses and analyses
$\{Z_{i, k}: i=1, \ldots, m ;\, k=1, \ldots, K\}$
from the assumed joint canonical distribution implied by the information fractions.\cite{jennison1999}
Nominal one-sided p-values are computed at each analysis by 
$p_{i, k}=1-\Phi(Z_{i, k})$, 
and the group sequential graphical procedure is applied forward through analyses
$k=1, \ldots, K$, recording $\boldsymbol{\tau}$
as the analysis times at which rejections are declared by the sequential procedure.
The expected gain is estimated by averaging $\psi(\mathbf{r}, \boldsymbol{\tau})$ across the simulated replicates.

\subsection{Elicitation of the gain function}\label{sec:methods_elicitation}

The gain function $\psi(\mathbf{r})$ encodes \textit{a priori} trial objectives, i.e., the relative desirability of different label-enabling rejection patterns -- rather than statistical operating characteristics.
The parameters of $\psi(\mathbf{r})$ as described in Equation~\eqref{eq:hurdle_formula} must therefore be elicited as value judgements from the trial-planning team, rather than determined from assumed effect sizes or correlations.
Such judgements should draw on external evidence that informs the \textit{value} of a claim-enabling rejection (not the probability of achieving it), for example, the commercial value of each claim or the competitor landscape.

We describe the elicitation of the gain function for a fixed-sample design for the case in which the incremental value of each hypothesis beyond the regulatory success criterion is additive.
(The extension to GSDs requires additionally specifying the discount function $d(\cdot)$, as illustrated in Example 5).
Let $\mathcal{P} \subseteq \{1, \ldots, m\}$ denote the set of hypotheses that define the regulatory success criterion, and let $\mathcal{S} = \{1, \ldots, m\} \setminus \mathcal{P}$ denote the remaining ``incremental'' hypotheses.
The regulatory success indicator $\mathbb{I}_{\text{reg}}(\mathbf{r})$ is any pre-specified Boolean function of $\{r_i : i \in \mathcal{P}\}$. The gain function takes the form
\begin{equation}\label{eq:additive}
\psi(\mathbf{r}) = \mathbb{I}_{\text{reg}}(\mathbf{r}) \cdot \biggl(v_{\text{base}} + \sum_{j \in \mathcal{S}} v_j r_j\biggr).
\end{equation}
The additivity assumption implicit in Equation~\eqref{eq:additive} --- that the value of rejecting $H_j$ ($j \in \mathcal{S}$) does not depend on which other hypotheses in $\mathcal{S}$ have been rejected --- keeps the number of parameters small and makes them easy to elicit.
Non-additive incremental components could also be accommodated within the general framework of Equation~\eqref{eq:hurdle_formula},
but would 
require a more complex elicitation than described here.

We propose a structured elicitation in which a cross-functional panel from the trial-planning team (e.g., clinical, commercial, and regulatory) independently rates the incremental value of each label claim conditional on regulatory success using a direct rating (DR) task on a numerical scale (e.g., 0-100).
The scale is anchored so that the maximum score is assigned to the most valuable secondary endpoint and other endpoints are scored as a percentage of that value.
This anchoring supports the ratio-scale interpretation required by the proportional allocation below.
DR is preferred over point allocation because it yields more reliable test-retest weights and imposes lower cognitive burden on respondents.\cite{bottomley2000}
Panellists score the incremental trial value of each confirmatory conclusion, not the probability of achieving it, and do so independently of other panellists. 
Results are then collected and presented back to all panellists in a group meeting --- with attention drawn to any large discrepancies --- after which panellists submit a final rating. These steps incorporate key elements of a modified Delphi process, namely independent rating, structured feedback, and re-rating, but in a deliberately lightweight form suited to rating a small number of incremental claim values.\cite{hasson2000} Finally, the aggregated score $\bar{s}_j$ for each $H_j$ ($j \in \mathcal{S}$) is taken as the median across panellists in the final round. 

The overall importance of key secondary conclusions relative to baseline regulatory success is elicited via an anchor question, such as: ``If achieving regulatory success alone has value 100, what is the value of regulatory success plus achieving all key label claims?''
Writing the consensus response as $S_{\text{all}}$, define $\kappa = (S_{\text{all}} - 100)/100$, so that $\kappa$ is the total incremental gain expressed as a fraction of baseline approval.
We allocate this gain across incremental hypotheses in proportion to the DR scores and normalise so that $\psi(\mathbf{r}) \in [0, 1]$:
\[
v_{\text{base}} = \frac{1}{1+\kappa}, \qquad v_j = \frac{\kappa}{1+\kappa} \cdot \frac{\bar{s}_j}{\sum_{\ell \in \mathcal{S}} \bar{s}_\ell}, \quad j \in \mathcal{S}.
\]
When regulatory success requires a single primary endpoint, or co-primary endpoints that must all be met, the optimal graph does not depend on $v_{\text{base}}$ and the anchor question can be skipped. 
The reason is that any graph is matched or improved by one that places full initial weight on the primary (or, for co-primaries, tests them first in fixed sequence at full $\alpha$) and recycles onward only once the hurdle is cleared.
The probability of regulatory success is then the same for every graph in this class, so $v_{\text{base}}$ contributes only a constant to the expected gain. The anchor question remains necessary when the trial values more than one route to approval and when the gain depends on rejection timing (see Examples 1 and 5 in Section~\ref{sec:examples}).

\paragraph*{Example of gain function elicitation for an osteoporosis trial}
\vspace*{6pt}

Returning to the osteoporosis trial of Section~\ref{sec:define_gain_function}: suppose the final-round consensus DR scores for the incremental rejections are $\bar s_2=90$ for $H_2$ and $\bar s_3=45$ for $H_3$, reflecting that avoiding hip fracture is judged twice as valuable as BMD conditional on primary success.
If the anchor question yields $S_{\mathrm{all}}=150$, then $\kappa=(150-100)/100=0.5$, meaning that achieving all key secondary conclusions together is valued at $50\%$ of baseline approval. Applying the mapping above gives
\[
v_{\text{base}}=\frac{1}{1+\kappa}=0.667,\qquad 
v_2=\frac{\kappa}{1+\kappa}\cdot\frac{90}{90+45}=0.222,\qquad
v_3=\frac{\kappa}{1+\kappa}\cdot\frac{45}{90+45}=0.111,
\]
and hence
\[
\psi(\mathbf r)=r_1\left(0.667+0.222\,r_2+0.111\,r_3\right).
\]
Thus primary success alone yields gain $0.667$, while achieving all key secondary conclusions yields the maximum gain of $1$, with $H_2$ contributing twice the incremental gain of $H_3$ conditional on primary success.

\subsection{Optimisation methods}

We write $\mathcal{U}(\mathbf{w}, \mathbf{G})$ to denote
a generic objective function of the graph. Depending on the context, 
$\mathcal U$ will represent either the conditional expected gain 
$U(\mathbf{w}, \mathbf{G} ; \boldsymbol{\theta})$ for a fixed scenario
$\boldsymbol{\theta}$, or the Bayes gain $U_B(\mathbf{w}, \mathbf{G})$ 
averaged over a prior on $\boldsymbol{\theta}$.

\subsubsection{Formulation of optimisation problem} \label{sec:formulation}

An optimal graphical procedure is defined by the weight vector $\mathbf{w}$ and 
the transition matrix $\mathbf{G}$ that maximises our chosen objective 
function $\mathcal{U}(\mathbf{w}, \mathbf{G})$ over a set of constraints
which define the search space; the constraints will always include the 
fundamental rules of graphical procedures and can be further restricted to 
reflect clinical or commercial priorities. 
The general form of the non-linear optimisation problem is to find the parameters 
$(\mathbf{w}^*, \mathbf{G}^*)$ for the graphical procedure that solves:
$$
(\mathbf{w}^*, \mathbf{G}^*) = \argmax_{(\mathbf{w}, \mathbf{G}) \in \Omega} \text{\ } \mathcal{U}(\mathbf{w}, \mathbf{G})
$$
where $\Omega$ is the feasible set defined by the constraints on the graph parameters, and optional additional constraints to yield desired properties in the graph, e.g., lower bounds on the \textit{local power} for specific hypotheses (the probability that the procedure rejects a hypothesis, accounting for the multiplicity adjustment). 

\paragraph*{Maximising expected gain for a fixed design}
\vspace*{6pt}

A confirmatory trial's sample size is often fixed in advance --- set by the power required for the primary endpoint(s), or by external operational and regulatory constraints --- and is then held fixed while the graph is optimised.
The optimisation then maximises either the
conditional expected gain at a fixed alternative $\boldsymbol{\theta}$, or the Bayes
gain integrated over the prior $\pi(\boldsymbol{\theta})$:
\begin{maxi}
{w_i, g_{ij}}{\mathcal{U}(\mathbf{w}, \mathbf{G})}{}{}\label{eq:fixed_opt_prob}
\addConstraint{0 \leq w_i \leq 1,}{\quad i = 1, \ldots, m}
\addConstraint{0 \leq g_{ij} \leq 1,}{\quad i,j = 1, \ldots, m \text{ and } i \neq j}
\addConstraint{g_{ii} = 0,}{\quad i=1, \dots, m}
\addConstraint{\sum_{i=1}^{m}w_i \leq 1}
\addConstraint{\sum_{j=1}^m g_{ij} \leq 1}{\quad \text{for } i = 1, \ldots, m.}
\end{maxi}
The constraints in the above optimisation problem are simply the regularity conditions for graphical procedures.
Additional constraints can be imposed on $\mathbf{G}$ to enforce a specific testing hierarchy, for example,
enforcing testing of primary hypotheses before secondary hypotheses by: 
(i) setting $w_j = 0$ for each secondary hypothesis $H_j$ (no initial weight on secondaries), and
(ii) setting certain transition weights $g_{ji} = 0$ to prevent alpha-recycling from secondary to primary endpoints.
A lower bound on the local power of a primary endpoint can be added as a further constraint where needed.

\paragraph*{Minimising sample size subject to constraint on conditional expected gain}
\vspace*{6pt}

Rather than optimising the graph at a fixed design, an alternative goal is to seek the smallest sample size at which some graph attains a target level of value.
Here, adequacy should be judged against a \textit{target product profile} (the regulatory hurdle together with the secondary claims the sponsor intends to pursue) rather than against primary power alone.
Assuming balanced allocation across arms, let $n$ denote the per-arm sample size, so that 
$N_{\text{total}} = A \cdot n$ for $A$ treatment arms, and assume the noncentrality parameter scales as 
$\Delta_i(n) = \delta_i \sqrt{n}$, where $\delta_i$
is the standardised effect size for the comparison associated with $H_i$.
We then seek the minimal $n$, solving the constrained problem:
\begin{mini}
{n, \mathbf{w}, \mathbf{G}}{n}{}{}\label{eq:sample_size_opt_prob}
\addConstraint{U(\mathbf{w}, \mathbf{G}; \boldsymbol{\Delta}(n), \boldsymbol{\Sigma}) \ge \gamma}
\addConstraint{\text{graph constraints on } (\mathbf{w}, \mathbf{G})}
\end{mini}
where $\gamma$ is the pre-specified target for the expected gain. e.g., $\gamma = 0.7$. 
The constraint on conditional expected gain can be replaced by desired lower bounds on local power for specific hypotheses; this approach is equivalent to the method described by Zhang et al. \cite{zhang2023}

\paragraph*{Group sequential optimisation problem}
\vspace*{6pt}

In the group sequential setting, the optimisation problem involves selecting the initial graph parameters ($\mathbf{w}, \mathbf{G}$) that maximise the conditional expected gain.
We treat the spending functions $\{\phi_1,...,\phi_m\}$ and analysis times  (and implied information fractions) as fixed inputs, since these are often fixed by operational or regulatory constraints.
The decision variables are the initial graph parameters. The optimisation problem for a target scenario $\boldsymbol{\theta}$ is:

\begin{maxi} 
{\mathbf{w}, \mathbf{G}}{\mathcal{U}(\mathbf{w}, \mathbf{G})}{}{}
\addConstraint{\text{graph constraints on } (\mathbf{w}, \mathbf{G})}
\end{maxi}
\\[1pt]
Note that the above problem can be readily extended to the joint optimisation of the graphical procedure and the GSD parameters.
If the calendar times of interim analyses $\mathbf{s}=(s_1,\dots,s_{K-1})$ and the spending function parameters are treated as decision variables rather than fixed inputs, the test statistic mean vector $\boldsymbol{\Delta}(s_k)$ and correlation matrix $\boldsymbol{\Sigma}(s_k)$ become functions of the analysis schedule through the rate of information accrual.
The time-dependent gain function $\psi(\mathbf{r},\boldsymbol{\tau})$ could then be used to find the optimal balance between the cost of delaying an analysis (discounted value via $d(\tau)$) and the benefit of waiting for more events (increased power as the noncentrality parameters grow).

\subsubsection{Computational approach}\label{sec:heuristic}

The decision-theoretic framework is agnostic to the optimiser: any method capable of handling constraints and a stochastic objective can be substituted.
Nevertheless, the optimisation problems in Section~\ref{sec:formulation} present specific computational challenges; the objective function is non-convex, may have optima on the boundary of the feasible set,\cite{zhan2022, xi2024} and must be estimated via Monte Carlo simulation.

We address these issues using two strategies. First, \textit{simulation sample reuse}: a single fixed matrix of $N_{\text{sim}}$ p-value vectors, simulated from $f(\mathbf{p} \mid \boldsymbol{\theta})$ (or $f^*$ for Bayes gain), is used to evaluate the objective for every candidate graph throughout the optimisation.\cite{wang2002}
This eliminates inter-iteration Monte Carlo noise, so that any observed change in the objective is attributable to the change in graph parameters.
Second, for the optimisation itself, we employ a hybrid evolutionary algorithm that combines a Fast Evolutionary Strategy for global exploration with intermittent derivative-free local search (Nelder--Mead) for refinement.\cite{yao1997, moscato1989}
The solution is a numerical approximation whose precision increases with $N_{\text{sim}}$ and the computational budget.
Finally, a pruning step tests each hypothesis and transition weight for removal, setting it to zero whenever doing so does not reduce the estimated gain, returning a simpler graph that achieves the same gain or higher.
Details of the algorithm, including adaptations for sample-size minimisation and GSDs, are given in the Supporting Information (Section~S2).

\section{Results: Applications in pharmaceutical trials}\label{sec:examples}

We illustrate the proposed framework through five examples drawn from real and hypothetical pharmaceutical trials, each highlighting a different aspect of the methodology.
All examples use a one-sided FWER of $\alpha = 0.025$ and adopt the multivariate normal test statistic model $\mathbf{Z} \sim \mathcal{N}(\boldsymbol{\Delta}, \boldsymbol{\Sigma})$, with the exception of Example 4, where uncertainty in the noncentrality parameters is incorporated through a prior on $\boldsymbol{\Delta}$. 
In each case, the noncentrality parameter for hypothesis $H_i$ is derived from the assumed nominal (unadjusted) power $1 - \beta_i$ at level $\alpha$ via 
\begin{equation}\label{eq:power_ncp}
\Delta_i = \Phi^{-1}(1-\alpha) + \Phi^{-1}(1-\beta_i).
\end{equation}
The objective function for each example is estimated by Monte Carlo simulation using $10^6$ replicates per iteration of the evolutionary algorithm and $5 \times 10^6$ replicates per iteration of the local search, with simulation sample reuse as described in Section~\ref{sec:heuristic}.
The evolutionary algorithm terminates after 1000 consecutive generations without improvement in the objective function.
The final local search terminates when every parameter changes by less than $5 \times 10^{-8}$ times its absolute value in a single step. Example 5 is an exception: it has a single free parameter, which was maximised on a dense grid rather than by the evolutionary algorithm (Supporting Information, Section~S2.3). The expected gains and local powers reported below were computed on a ``fresh'' independent sample of $5 \times 10^6$ p-value vectors, drawn after optimisation was complete.

\subsection{Example 1: Clinical trial with an enriched subgroup}
\label{sec:app_subgroup_total}

We consider a randomised, placebo-controlled phase~III trial with two pre-specified analysis populations: an enriched subgroup $E$ comprising two-fifths of the total population, and the total population $T$. 
In each population, three endpoints are considered: one primary endpoint and two key secondary endpoints. This yields the family of $m=6$ null hypotheses
$
\mathcal{H}=\{H_{E,1},H_{E,2},H_{E,3},H_{T,1},H_{T,2},H_{T,3}\},
$
where the numerical index $j\in\{1,2,3\}$ denotes the primary, secondary~1, and secondary~2 endpoints, respectively.

\begin{table}[!ht]
\centering
\caption{Assumed one-sided per-hypothesis nominal (unadjusted for multiplicity) powers \(1 - \beta_{P,j}\) and corresponding noncentrality parameters $\Delta_{P,j}$.}
\label{tab:subgroup_inputs}
\begin{tabular}{llcc}
\toprule
\textbf{Population} & \textbf{Hypothesis} & \textbf{Nominal Power} $(1 - \beta_{P,j})$ & \textbf{Noncentrality} $(\Delta_{P,j})$ \\
\midrule
\multirow{3}{*}{Enriched (\(E\))} 
& $H_{E,1}$ (primary)      & $0.90$ & $3.24$ \\
& $H_{E,2}$ (secondary 1)  & $0.90$ & $3.24$ \\
& $H_{E,3}$ (secondary 2)  & $0.86$ & $3.04$ \\
\midrule
\multirow{3}{*}{Total ($T$)} 
& $H_{T,1}$ (primary)      & $0.88$ & $3.13$ \\
& $H_{T,2}$ (secondary 1)  & $0.88$ & $3.13$ \\
& $H_{T,3}$ (secondary 2)  & $0.84$ & $2.95$ \\
\bottomrule
\end{tabular}
\end{table}

The assumed per-hypothesis nominal powers and the resulting noncentrality parameters --- derived via Equation~\eqref{eq:power_ncp} --- are shown in Table~\ref{tab:subgroup_inputs}.
Despite the smaller sample size in the enriched subgroup, the larger assumed treatment effect yields higher nominal power for the enriched hypotheses than for their total-population counterparts.
We assume a compound-symmetry structure within each population, so that for $j\neq \ell$, we have
$
\operatorname{Corr}(Z_{P,j},Z_{P,\ell})=\frac{1}{2}
$, where $P\in\{E,T\}$.
Because the enriched population is nested within the total population, the test statistics for the \emph{same} endpoint across populations are correlated. With the enriched subgroup prevalence being two-fifths of the total, then 
$
\operatorname{Corr}(Z_{E,j},Z_{T,j})=\sqrt{\frac{2}{5}}$ for $j\in\{1,2,3\}$
and for $j\neq \ell$ we take the cross-population, cross-endpoint correlation as the product 
$
\operatorname{Corr}(Z_{E,j},Z_{T,\ell})=\frac{1}{2} \times \sqrt{\frac{2}{5}}.
$

\begin{figure}
\centerline{\includegraphics[width=120mm]{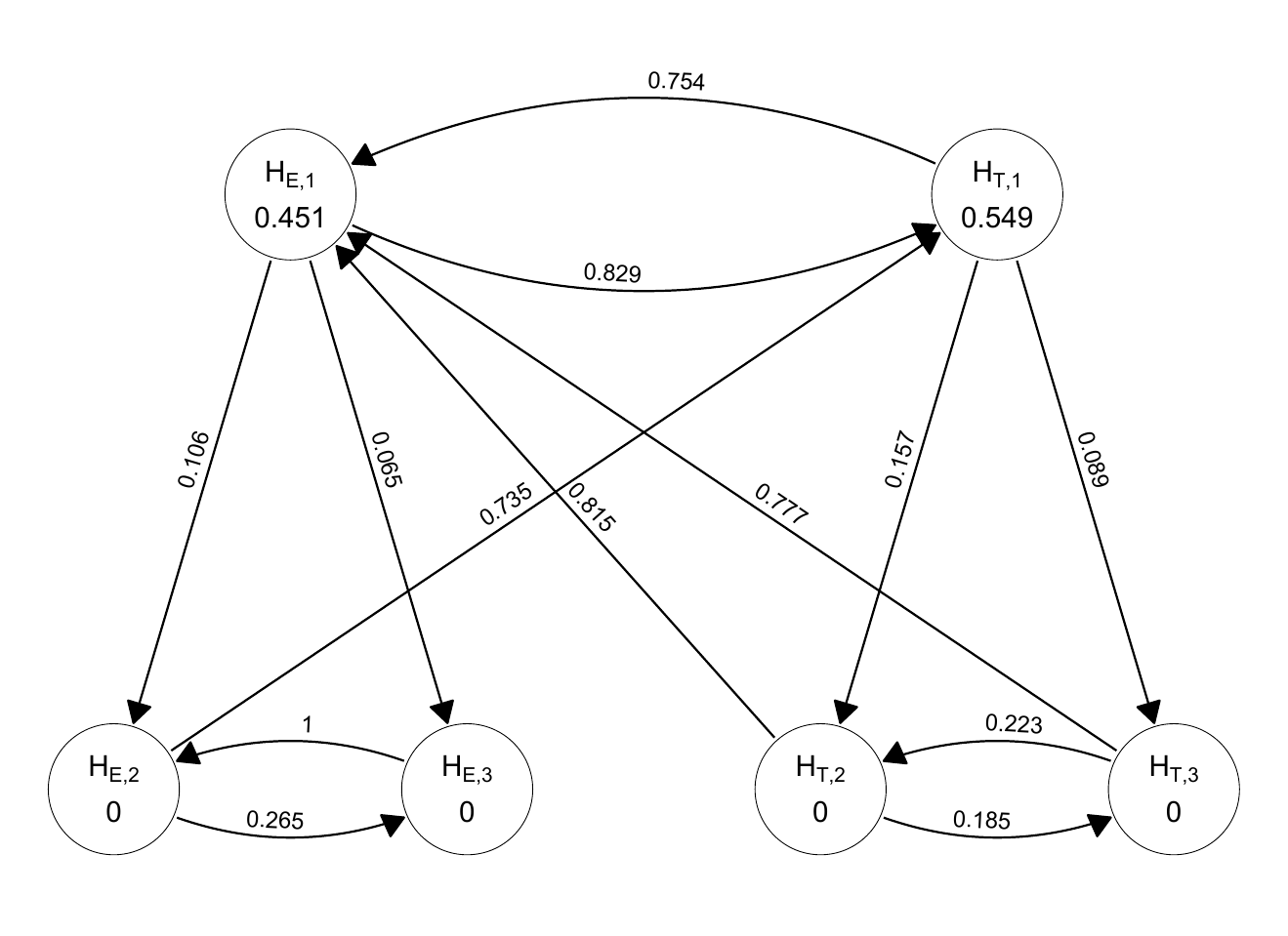}}
\caption{Optimised graphical procedure for Example 1. The subsets $E$ and $T$ represent the enriched and total population hypotheses respectively. \label{fig:subgroup_graph}}
\end{figure}

For this case study, we assume that the trial-planning team agreed in advance on how to value the possible outcomes. A treatment approved for use in all patients (the total population) was judged more valuable than one approved only for the biomarker-positive subgroup, because it would be available to a larger number of patients.
However, approval in the subgroup alone was still considered valuable because of the greater severity of disease in these patients.\cite{dmitrienko2017} 
Within each population, key secondary endpoints were agreed to contribute one quarter of the value of that population's primary claim, since each supports a further label claim that adds clinical benefit and commercial value, but only once that population's primary claim is secured.

Let $\mathbf{r} = (r_{E,1}, r_{E,2}, r_{E,3}, r_{T,1}, r_{T,2}, r_{T,3})$ denote the rejection pattern.
We assign baseline values $v_T~=~4$ and $v_E~=~3$ to primary-endpoint rejection in the total and enriched populations respectively, reflecting the greater commercial value of a broad label claim.
Each key secondary endpoint contributes one quarter of the baseline value of its own population, conditional on that population's primary success: i.e., 
$v_{T,2} = v_{T,3} = v_T\,/\,4 = 1$ and $v_{E,2} = v_{E,3} = v_E\,/\,4 = 0.75$.
The (unnormalised) gain function is
\[
\psi(\mathbf{r}) = r_{T,1}(v_T + v_{T,2}\,r_{T,2} + v_{T,3}\,r_{T,3}) + r_{E,1}(v_E + v_{E,2}\,r_{E,2} + v_{E,3}\,r_{E,3}),
\]
or equivalently, substituting the numerical values,
\[
\psi(\mathbf{r}) = r_{T,1}(4 + r_{T,2} + r_{T,3}) + r_{E,1}(3 + 0.75\,r_{E,2} + 0.75\,r_{E,3}).
\]
Thus $\psi$ accrues value through two population-specific gates: a broad claim ($r_{T,1} = 1$), a restricted claim ($r_{E,1} = 1$), or both, with secondary endpoints contributing only within populations that achieve primary success.
The two population claims are treated as additive, so the enriched-population claim retains its value even when the broad claim is secured; this reflects the greater disease severity in biomarker-positive patients, for whom a subgroup-specific claim might add value alongside a broad approval.\cite{dmitrienko2017}
If the subgroup claim were instead judged redundant once the broad label is won, its contribution can be replaced with $r_{E,1}(1 - r_{T,1})(v_E + \cdots)$.
The maximum is $\psi_{\max} = 4 + 2 + 3 + 1.5 = 10.5$; for interpretability we report the normalised gain, $\psi(\mathbf{r})/\psi_{\max} \in [0,1]$.

The graph optimisation had a constrained transition matrix:
\[
\mathbf{G} = \begin{pmatrix} 
0 & g_{E1,E2} & g_{E1,E3} & g_{E1,T1} & 0 & 0 \\ 
0 & 0 & g_{E2,E3} & g_{E2,T1} & 0 & 0 \\
0 & g_{E3,E2} & 0 & g_{E3,T1} & 0 & 0 \\
g_{T1,E1} & 0 & 0 & 0 & g_{T1,T2} & g_{T1,T3} \\ 
g_{T2,E1} & 0 & 0 & 0 & 0 & g_{T2,T3} \\
g_{T3,E1} & 0 & 0 & 0 & g_{T3,T2} & 0 \end{pmatrix}
\]
where entries marked $g_{..}$ are free decision variables and all zero entries are fixed.
Alpha-recycling occurs from each primary to its own secondary endpoints and to the other primary; secondary endpoints alpha-recycle within their population and to the other population's primary, but not back to their own primary.
Solving the optimisation problem~\eqref{eq:fixed_opt_prob} to find the graph under these constraints that maximises the conditional expected gain 
$U(\mathbf{w},\mathbf{G};\boldsymbol{\Delta},\boldsymbol{\Sigma})$ yields the graph in Figure~\ref{fig:subgroup_graph}. 
Table~\ref{tab:subgroup_results} compares the resulting optimised graph to (i) a fixed-sequence strategy (in descending order of nominal powers) and (ii) a Holm-Bonferroni procedure.

\begin{table}[!ht]
\centering
\caption{Performance comparison for Example 1 under the assumed $(\boldsymbol{\Delta},\boldsymbol{\Sigma})$: conditional expected gain and local (adjusted for multiplicity) powers under compared approaches.}
\label{tab:subgroup_results}
\begin{tabular}{lccccccc}
\toprule
& \textbf{Conditional} & \multicolumn{6}{c}{\textbf{Local power}} \\
\cmidrule(lr){3-8}
\textbf{Procedure} & \textbf{expected gain} & $H_{E,1}$ & $H_{E,2}$ & $H_{E,3}$ & $H_{T,1}$ & $H_{T,2}$ & $H_{T,3}$ \\
\midrule
Optimised graph (Figure \ref{fig:subgroup_graph}) & \textbf{0.814} & 0.871 & 0.748 & 0.698 & 0.858 & 0.735 & 0.688
\\
Fixed sequence & 0.778 & 0.900 & 0.827 & 0.645 & 0.763 & 0.713 & 0.597 \\
Holm-Bonferroni  &0.770 & 0.819 & 0.819 & 0.772 & 0.796 & 0.796 & 0.751 \\
\bottomrule
\end{tabular}
\end{table}

The optimised graph achieves a conditional expected gain of 0.814, a relative improvement of 5\% over the fixed-sequence (0.778) and 6\% over the Holm (0.770) procedure.

\subsection{Example 2: Sample size optimisation to achieve a submission criterion (success for at least one dose)}
\label{sec:app_dose_success}

The example is motivated by the FIBRONEER-ILD phase~III trial, which evaluates two doses (High and Low) of a novel PDE4B inhibitor in patients with interstitial lung disease. \cite{maher2025}
Patients were randomised in a 1:1:1 ratio to receive nerandomilast at a dose of 18 mg twice daily, nerandomilast at a dose of 9 mg twice daily, or placebo.
We adapt the design to define two efficacy objectives: a physiological primary endpoint (continuous FVC) and a clinical secondary endpoint (binary composite failure: exacerbation, hospitalisation, or death). 
This yields the family of $m = 4$ null hypotheses: $\mathcal{H} = \{H_1, H_2, H_3, H_4\}$ where the indices are grouped by dose level:
$H_1$ and $H_2$ correspond to the High Dose (primary and secondary endpoints), while $H_3$ and $H_4$ correspond to the Low Dose.

When two doses are carried into a confirmatory trial, the sponsor is more likely to seek approval for a single efficacious dose than for both simultaneously.
If both doses demonstrate similar efficacy, the lower dose is generally preferred for the label because it is expected to offer a more favourable benefit-risk profile.
Pursuing a single dose also avoids the additional manufacturing and supply-chain complexity of supporting two dose strengths.
The gain function should therefore reward dual-endpoint success for \emph{at least one} dose, without requiring that both doses succeed.
We therefore apply a minimal
submission ``win criterion'' gain function that assigns unit gain if at least one dose meets \emph{both} endpoints, and zero otherwise:
\begin{equation}
\psi(\mathbf{r})
=
\mathbb{I}\left\{r_1 r_2 + r_3 r_4 \ge 1\right\},
\label{eq:dose_win_criterion}
\end{equation}
where $\mathbf{r}=(r_1,r_2,r_3,r_4)$ denotes the rejection pattern induced by the graphical procedure. Note that the gain function \eqref{eq:dose_win_criterion} has no incremental component $\psi_{\text {incr}}(\mathbf{r})$, and neither does \eqref{eq:dose_win_criterion} assign additional value if the trial simultaneously achieved dual
success at both doses. (Of course, in practice, a sponsor may value both doses being approved; we assume this is not the case for this example).

For illustration the continuous FVC endpoint is modelled with standardised treatment effects of $\delta_1=0.250$ and $\delta_3=0.230$; the binary composite endpoint has lower standardised treatment effect for both doses ($\delta_2=0.150$ and $\delta_4=0.145$).
Consistent with FIBRONEER-ILD protocol assumptions, we model a correlation of $\rho=0.4$ between endpoints within the same dose, and $\rho=0.5$ between doses for the same endpoint, yielding the correlation structure:
\[
\boldsymbol{\Sigma}=\left(\begin{array}{cccc}
 1.0 & 0.4 & 0.5 & 0.2 \\
0.4 & 1.0 & 0.2 & 0.5 \\
 0.5 & 0.2 & 1.0 & 0.4 \\
 0.2 & 0.5 & 0.4 & 1.0
\end{array}\right).
\]
We define our goal to be to optimise the graph so as to find the minimum sample size to achieve a target conditional expected gain to be 85\% of the maximum, 
i.e., a probability of achieving dual-endpoint success for at least one dose of 85\%, conditional on the design assumption 
$\boldsymbol{\theta}=(\boldsymbol{\Delta}(n), \boldsymbol{\Sigma})$, where $\Delta_i(n)=\delta_i \sqrt{n}$. 
This is achieved by solving the optimisation problem defined in \eqref{eq:sample_size_opt_prob}.

\begin{figure}
\centerline{\includegraphics[width=120mm]{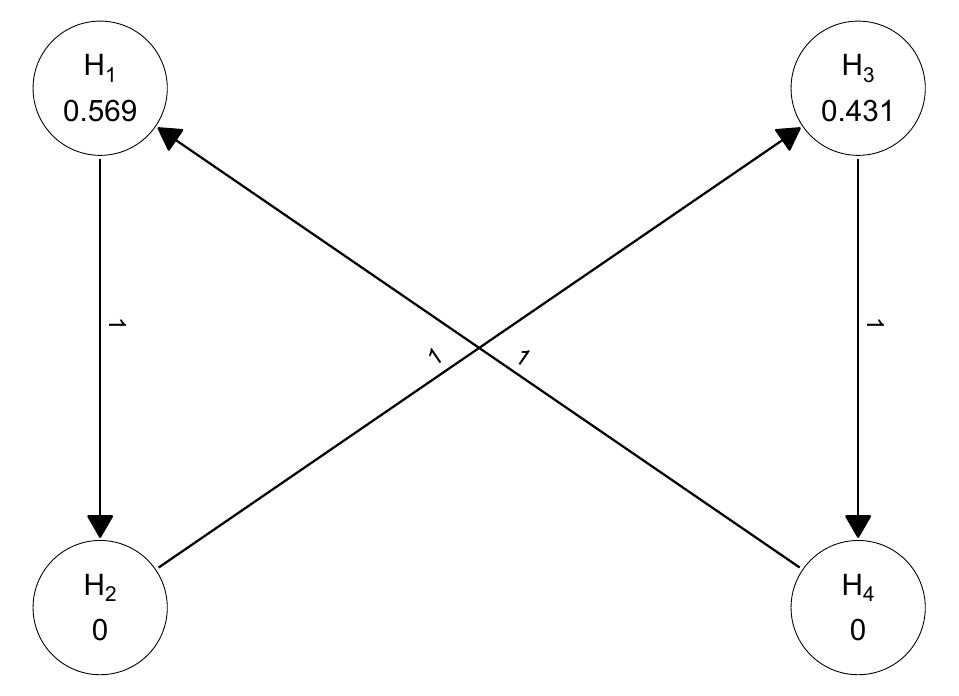}}
\caption{Optimised graphical procedure for Example 2, when $n = 360$ targeting a conditional expected gain of $0.85$ under the design assumptions described in Section~\ref{sec:app_dose_success}. 
\label{fig:sample_size_fig}}
\end{figure}

The resulting optimised group size is $n = 360$, corresponding to a total sample size $N_{\text{total}} = 1080$ (assuming balanced allocation), which is achieved using the graph displayed in Figure~\ref{fig:sample_size_fig}.
For comparison, the optimised graphical procedure achieves the target conditional expected gain of 85\% with 120 fewer participants than the best performing fixed sequence procedure (testing order $H_1 \to H_2 \to H_3 \to H_4$, i.e., testing High Dose endpoints before Low Dose), which achieves this with $N_{\text{total}} = 1200$.
A Holm-Bonferroni procedure is close to optimal under these assumptions, achieving the target expected gain at $N_{\text{total}} = 1086$.

\subsection{Example 3: Elicitation of incremental gain for key secondary endpoints}

We base this example on the ECZTRA phase~III programme evaluating tralokinumab in combination with corticosteroids compared with placebo control in patients with
moderate-to-severe atopic dermatitis.\cite{silverberg2021}
The trials used a number of endpoints to evaluate severity 16 weeks post-randomisation. Two were co-primary:
Investigator's Global Assessment (IGA) 0/1 and Eczema Area and Severity Index-75 (EASI-75). One is a key secondary endpoint, SCORing Atopic Dermatitis (SCORAD). Other key secondary endpoints 
adjusted for FWER control included a clinically meaningful improvement 
in pruritus, change in Dermatology Life 
Quality Index (DLQI), both also evaluated at week 16. 
Limiting the design to these endpoints yields the family of $m = 5$ null hypotheses: $\mathcal{H} = \{H_1, H_2, H_3, H_4, H_5\}$, displayed with illustrative design inputs in Table~\ref{tab:elicitation_power_inputs}.
We assume the correlation matrix for the corresponding test statistics:
\[
\boldsymbol{\Sigma}=\left(\begin{array}{ccccc}
1 & 0.8 & 0.5 & 0.5 & 0.8 \\
0.8 & 1 & 0.5 & 0.5 & 0.8 \\
0.5 & 0.5 & 1 & 0.5 & 0.5 \\
0.5 & 0.5 & 0.5 & 1 & 0.5 \\
0.8 & 0.8 & 0.5 & 0.5 & 1
\end{array}\right),
\]
reflecting that clinician-reported severity (IGA, EASI, SCORAD) are likely strongly
correlated, and that patient-reported outcomes (Pruritus, DLQI) correlate more moderately with disease severity endpoints and with each other.

\begin{table}[!ht]
\centering
\caption{Assumed one-sided per-hypothesis nominal (unadjusted for multiplicity) powers \(1 - \beta_{j}\) and corresponding noncentrality parameters $\Delta_{j}$ for Example 3.}
\label{tab:elicitation_power_inputs}
\begin{tabular}{llcc}
\toprule
\textbf{Hypothesis} & \textbf{Endpoint} & \textbf{Nominal power} $1-\beta_i$ & $\Delta_i$ \\
\midrule
$H_1$ & IGA 0/1 & 0.90 & 3.24 \\
$H_2$ & EASI-75 & 0.90 & 3.24 \\
$H_3$ & Pruritus & 0.80 & 2.80 \\
$H_4$ & DLQI & 0.70 & 2.48 \\
$H_5$ & SCORAD & 0.85 & 3.00 \\
\bottomrule
\end{tabular}
\end{table}

Regulatory success requires rejection of both co-primaries ($\mathbb{I}_{\text {reg}}(\mathbf{r}) = r_1\,r_2$). Assuming additive incremental value for each secondary rejection:
\[
\psi(\mathbf{r})=r_1 r_2\left(v_{\text{base}}+v_3 r_3+v_4 r_4+v_5 r_5\right).
\]
where the incremental gain, $\psi_\text{incr}=v_3 r_3+v_4 r_4+v_5 r_5$.
Suppose a cross-functional panel (clinical, commercial, regulatory strategy) independently rated the incremental value of each secondary rejection 
conditional on co-primary success, on a $0-100$ scale, as described in Section~\ref{sec:methods_elicitation}. 
For illustration, suppose the final-round median DR scores are 85 for pruritus, 50 for DLQI and 15 for SCORAD, so pruritus is judged just over five times as valuable as SCORAD conditional on co-primary success. These sum to 150, giving relative weights of 0.567, 0.333 and 0.100.
The panel was asked the anchor question: ``If achieving co-primary success has value 100, what is the value of achieving co-primary
success plus meeting all three key secondary endpoints?'' The consensus response was $S_{\text{all}}=180$. This implies a total incremental value of $80\%$ of
the baseline, i.e. $\kappa=(180-100) / 100=0.8$. So that the gain is in $[0,1]$ with maximum achievable gain equal to 1, we set:
\[
v_{\text{base}}=\frac{1}{1+\kappa}=\frac{1}{1.8}=0.556, \quad\left(v_3, v_4, v_5\right)=\frac{\kappa}{1+\kappa} \cdot \left(0.567,0.333, 0.100\right)=(0.252,0.148,0.044).
\]
This yields the conditional gain function:
\[
\psi(\mathbf{r})=r_1 r_2\left(0.556+0.252\, r_3+0.148\, r_4+0.044\, r_5\right).
\]

We constrain the graphical procedure so that the co-primary endpoints (IGA, EASI) are tested first in fixed sequence. 
On rejection of both $H_1$ and $H_2$, alpha-recycling proceeds among the secondary endpoints. Therefore the transition weights $g_{ij}$ for
$i \in\{3,4,5\}$ and $j \in\{3,4,5\}$, $j \neq i$, are free decision variables.
We also fix $g_{i 1}=g_{i 2}=0$ for all $i \in\{3,4,5\}$, preventing
any recycling back to $\left\{H_1, H_2\right\}$.

\begin{figure}
\centerline{\includegraphics[width=120mm]{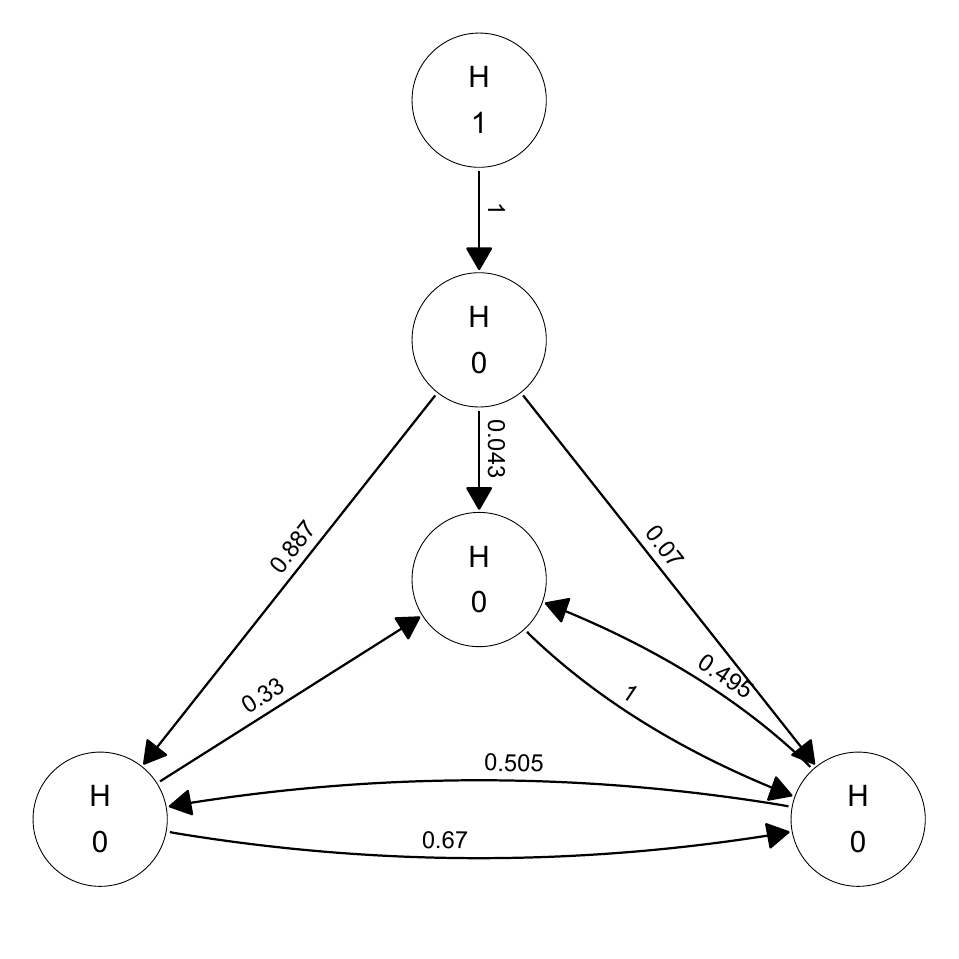}}
\caption{
Optimised graphical procedure for Example 3.
\label{fig:elicit_graph}
}
\end{figure}

We solve the optimisation problem \eqref{eq:fixed_opt_prob}, maximising the conditional expected gain $U(\mathbf{w}, \mathbf{G}; \boldsymbol{\theta})$. 
The resulting optimised graphical procedure is shown in Figure~\ref{fig:elicit_graph}, and Table~\ref{tab:elicit_results}
summarises its performance under the assumed $\boldsymbol\theta = (\boldsymbol{\Delta}, \boldsymbol{\Sigma})$. For comparison, we also evaluate three constrained testing strategies that preserve the co-primary
gatekeeping $H_1 \rightarrow H_2$:
(i) a value-ordered fixed sequence based on the elicited incremental weights (highest to lowest); 
(ii) a fixed sequence descending in nominal power (i.e., decreasing $\Delta_i$); and
(iii) a standard Holm procedure applied to $\left\{H_3, H_4, H_5\right\}$.
Table~\ref{tab:elicit_results} summarises the resulting conditional expected gains and local powers under the same inputs as the optimised graph.

\begin{table}[!ht]
\centering
\caption{Performance comparison for Example 3 under the assumed joint distribution $\boldsymbol{\theta}=(\boldsymbol{\Delta},\boldsymbol{\Sigma})$. The conditional expected gain is $U=\mathbb{E}\left[\psi(\mathbf r)\right]$. Local powers are reported for $H_3$--$H_5$.}
\label{tab:elicit_results}

\begin{minipage}{0.95\linewidth}
\centering
\begin{tabular}{lcccc}
\toprule
& & \multicolumn{3}{c}{\textbf{Local power}} \\
\cmidrule(lr){3-5}
\textbf{Procedure} & \textbf{Conditional expected gain*} & $H_3$ & $H_4$ & $H_5$ \\
\midrule
Optimised graph (Figure \ref{fig:elicit_graph}) & \textbf{0.774} & 0.717 & 0.580 & 0.721 \\
Value-ordered sequence ($H_1 \to H_2 \to H_3 \to H_4 \to H_5$) & 0.769 & 0.726 & 0.579 & 0.558 \\
Power-ordered sequence ($H_1 \to H_2 \to H_5 \to H_3 \to H_4$) & 0.767 & 0.689 & 0.558 & 0.799 \\
$H_1 \to H_2 \to \text{Holm}\{H_3, H_4, H_5 \}$ & 0.770 & 0.679 & 0.607 & 0.749 \\
\bottomrule
\end{tabular}

\vspace{4pt}
\footnotesize $^{*}$Gain values are scaled to lie in $[0,1]$.
\end{minipage}
\end{table}

The value-ordered and power-ordered fixed sequences illustrate the tension between prioritising the most valuable secondary endpoint (pruritus)
and the one most likely to reach significance (SCORAD).
The optimised graph achieves the highest expected gain (0.774) by balancing this trade-off via recycling among 
the secondary endpoints rather than committing to a single pre-specified order. 
The absolute improvement in expected gain is modest (0.004 over Holm).
However, for a treatment with projected annual revenues in the hundreds of millions, even small differences in expected gain can represent substantial economic value.\cite{lisovskaja2015}
The gain function makes this trade-off explicit and auditable.

\subsection{Example 4: Optimising Bayes gain under uncertainty in treatment effects}

This example illustrates how the optimised graphical procedure can differ when
performance is evaluated at a single fixed alternative (conditional expected gain)
versus performance averaged over design-stage uncertainty (Bayes gain).
We consider a confirmatory family of $m=4$ one-sided hypotheses 
$
\left\{H_1, H_2, H_3, H_4\right\}
$, 
with $H_1$ being a primary endpoint and $H_{2}$--$H_{4}$ key secondary endpoints. The gain function is
\[
\psi(\mathbf{r})=r_{1}\left(0.4+0.2r_{2}+0.2r_{3}+0.2r_{4}\right),
\]
so that no gain is accrued unless $H_{1}$ is rejected; rejection of the primary alone yields gain $0.4$,
and each secondary rejection contributes $0.2$ conditional on primary success. We condition on $\boldsymbol \Delta$ as described below, retaining the fixed correlation $\boldsymbol \Sigma$
\[
\mathbf{Z} \mid \boldsymbol \Delta \sim\mathcal{N}(\boldsymbol{\Delta},\boldsymbol{\Sigma}).
\]
For this example, we treat the correlation structure $\boldsymbol{\Sigma}$ as a fixed design-input reflecting patient-level dependence in the endpoints; specifically, we use a compound-symmetry structure
$\Sigma_{ij}=\rho$ for $i\neq j$, taking $\rho=0.7$. 

First, the graph is optimised under a fixed scenario $\boldsymbol{\theta}_{0}=(\boldsymbol{\Delta},\boldsymbol{\Sigma})$
by maximising the conditional expected gain $U(\mathbf{w},\mathbf{G};\boldsymbol{\theta}_{0})$
(Section~\ref{sec:expected_gain}). The noncentrality vector is $\boldsymbol \Delta = (3.24, 3.24, 2.80, 2.48)$,
which corresponds to nominal (unadjusted) powers of $(0.90,0.90,0.80,0.70)$
at one-sided level $\alpha = 0.025$.
Solving \eqref{eq:fixed_opt_prob} under these inputs yields
a fixed-sequence optimal graph ($H_1 \to H_2 \to H_3 \to H_4$), with performance summarised in Table~\ref{tab:bayes_results}.

\begin{figure}[ht]
    \centering
    
    \begin{subfigure}[b]{0.49\textwidth}
        \centering
        \includegraphics[width=\linewidth]{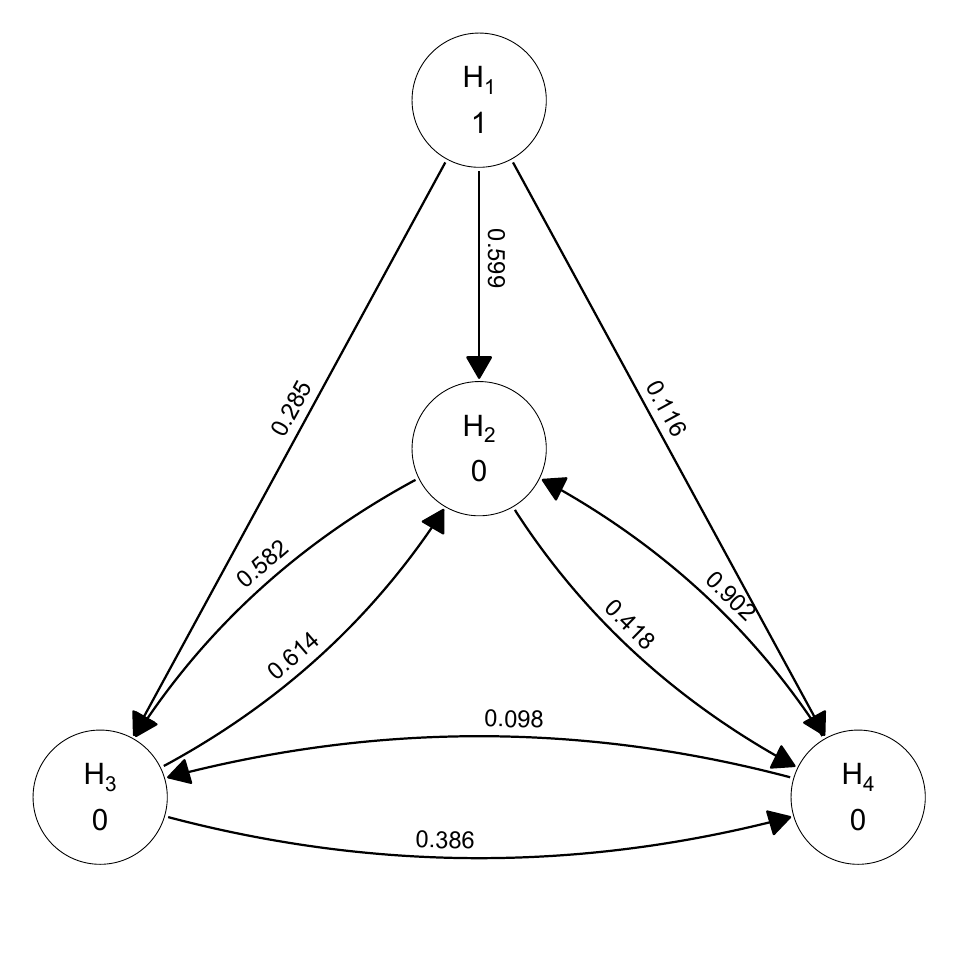}
        \caption{Bayes-optimal graphical procedure under the prior standard deviation $\sigma_{\Delta,i}=0.8$.}
        \label{fig:bayes_optimal_graph}
    \end{subfigure}
    \hfill
    \begin{subfigure}[b]{0.49\textwidth}
        \centering
        \includegraphics[width=\linewidth]{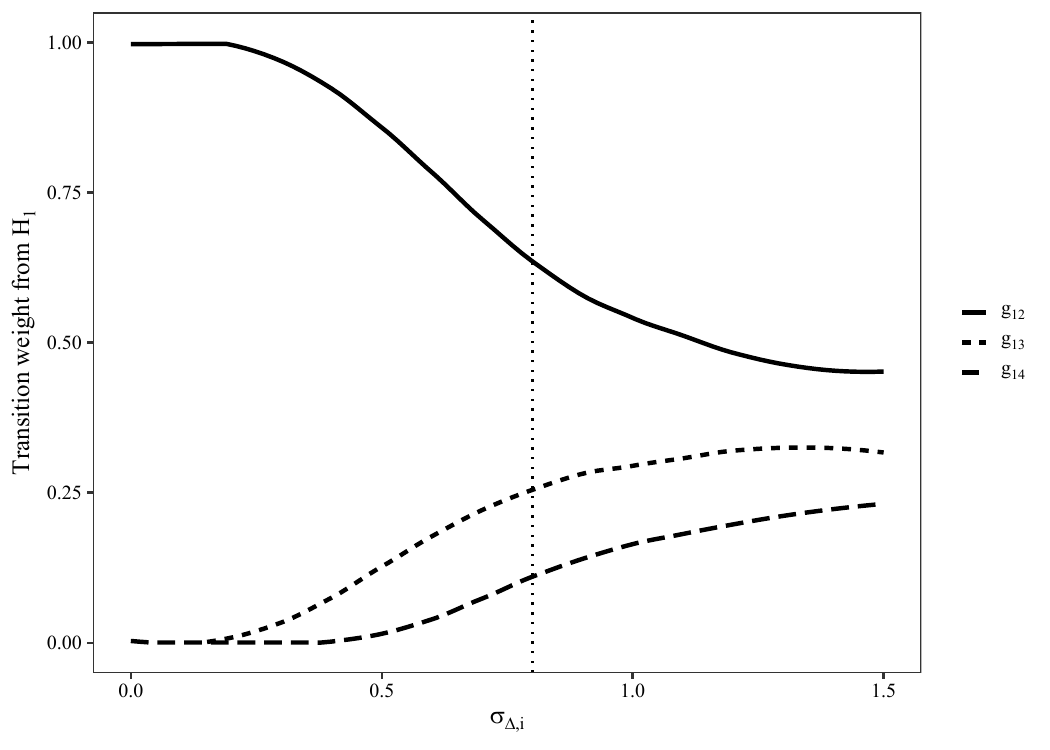}
        \caption{Bayes-optimal primary-to-secondary recycling split $(g_{12}, g_{13}, g_{14})$ as a function of $\sigma_{\Delta,i} \in[0,1.5]$. The vertical dotted line illustrates $\sigma_{\Delta,i} = 0.8$, i.e., the graph in (a).}  
        \label{fig:tau_sensitivity}
    \end{subfigure}
    
    \caption{Bayes-optimal graph and sensitivity to treatment effect prior standard deviation, $\sigma_{\Delta,i}$, in key secondary hypotheses $H_i$ ($i=2,3,4$).}
    \label{fig:combined_plot}
\end{figure}

\begin{table}[!h]
\centering
\caption{Conditional expected gain under a fixed alternative versus Bayes gain under design-stage uncertainty for Example 4.}
\label{tab:bayes_results}
\begin{tabular}{llcc}
\toprule
\textbf{Evaluation scenario} & \textbf{Objective function}
& \textbf{Fixed-sequence} 
& \textbf{Bayes-optimal graph} \\
& (type of gain) & (optimal for fixed $\boldsymbol{\theta}_0$) 
& (optimal for $\sigma_\Delta=0.8$) \\
\midrule
Fixed alternative $\boldsymbol{\theta}_0$ 
& $U(\mathbf{w},\mathbf{G};\boldsymbol{\theta}_0)$ 
& 0.803 
& 0.796 \\

Bayes prior ($\sigma_\Delta=0.8$) 
& $U_B(\mathbf{w},\mathbf{G})$ 
& 0.742 
& 0.759 \\
\bottomrule
\end{tabular}

\vspace{2mm}
\footnotesize
Reported gains are scaled to lie in $[0,1]$.
\end{table}
Second, we incorporate uncertainty in treatment effects by placing a prior on $\boldsymbol{\Delta}$.
Specifically, we assume independent normal priors
\[
\Delta_{i}\sim\mathcal{N}(\mu_{i},\sigma_{\Delta,i}^{2}),
\]
centred at the same mean vector $\mu$ implied by the nominal powers above. We fix
$\sigma_{\Delta,1}=0.2$ for the primary endpoint and set $\sigma_{\Delta,2}=\sigma_{\Delta,3}=\sigma_{\Delta,4}=\sigma_\Delta$ for the
key secondary endpoints, reflecting greater design-stage uncertainty for secondary
effects. For each $\sigma_\Delta$, we estimate Bayes gain $U_B(\mathbf{w}, \mathbf{G})$ by
simulation from the predictive distribution for the p-values induced by the prior on
$\boldsymbol{\Delta}$ together with the fixed within-trial correlation
$\boldsymbol{\Sigma}$, and re-optimise to obtain the Bayes-optimal graph.
The optimal graph for the value $\sigma_{\Delta,2}=\sigma_{\Delta,3}=\sigma_{\Delta,4}=0.8$ is shown in
Figure~\ref{fig:bayes_optimal_graph}, with performance also
summarised in Table~\ref{tab:bayes_results}.

To visualise how uncertainty in key secondary hypothesis treatment effects influences the
optimal graph, we vary $\sigma_\Delta$ and record the primary-to-secondary alpha-recycling split $\left(g_{12}, g_{13},
g_{14}\right)$. 
Figure~\ref{fig:tau_sensitivity} shows that as $\sigma_\Delta$ increases, it leads to
greater recycling across key secondary endpoints as the true order of treatment effect magnitude in endpoints becomes more uncertain. 
Note that the prior used on $\boldsymbol{\Delta}$ assumes independence between endpoints;
the result being that predictive correlation between the test statistics are attenuated as $\sigma_\Delta$ increases.
In practice, treatment effects on related endpoints are likely positively associated: a drug that works well on one measure of disease severity probably works well on another. 
A joint prior on $\boldsymbol{\Delta}$
with positive off-diagonal covariance would maintain stronger predictive correlations even at large $\sigma_\Delta$,
and would likely produce less recycling than observed in this example.
These independent-prior results for this illustrative example  should therefore be interpreted as an upper bound on the degree of alpha recycling induced at this level of marginal prior uncertainty. 
In general, the prior on $\boldsymbol{\Delta}$ should reflect genuine consensus beliefs about treatment effects at the design-stage -- informed by early-phase data and related compounds -- rather than the fixed point alternative used in conventional power calculations.
This is analogous to using a ``design prior'' for assurance as opposed to a single alternative used for a frequentist power calculation.\cite{hagan2005}
Joint priors and priors on the correlation matrix are discussed in the Section~S1 of the Supporting Information.

\subsection{Example 5: Optimising a graphical procedure for a group sequential design using time-dependent gain function}\label{sec:gsd_example}

We consider a confirmatory family $\mathcal{H}=\{H_{\text{PFS}},H_{\text{OS}}\}$ corresponding to progression-free survival (PFS, $H_{\text{PFS}}$) and overall survival (OS, $H_{\text{OS}}$) endpoints.
This example is drawn from oncology trials in which PFS matures earlier than OS, such as the phase~III EV-302 trial comparing the efficacy of enfortumab vedotin combined with pembrolizumab with platinum-based chemotherapy in patients with previously untreated locally advanced or metastatic urothelial carcinoma.\cite{powles2024, ema_padcev_2024}
In many oncology programmes, the practical value of a statistically significant result depends not only on which endpoint is positive, but also when the evidence becomes available.
Whilst regulators often consider OS to be the most important (and therefore valuable) primary endpoint, earlier endpoints such as PFS can still provide incremental value by supporting earlier evidence generation and physician confidence while OS matures, or even be used for submitting accelerated approval applications.\cite{FDA2025} Furthermore, the practical values of OS and earlier endpoints will vary between disease types and histology; trials in indolent haematological malignancies or rare cancers for which expected survival is relatively long would convey little value for improvements in OS and higher value for improvements in earlier endpoints such as PFS and Overall Response Rate (ORR).\cite{merino2023} 

As in the EV-302 design, we consider a trial that is monitored at $K=2$ analyses: an interim analysis at OS information fraction $t_{\text{OS},1} = 0.7$ (calendar time $s_1 = 37$ months) and a final analysis at $t_{\text{OS},2} = 1$ ($s_2 = 47$ months).
The PFS test statistic is treated as final at the interim analysis and unchanged thereafter, i.e., $t_{\text{PFS},1}=t_{\text{PFS},2} = 1$.
PFS is frozen from the first analysis, so its information fraction is already 1 at the interim and the same nominal p-value applies at both analyses.
Thus, a single one-sided PFS p-value
$
p_{\text{PFS}}=1-\Phi\left(Z_{\text{\text{PFS}}}\right)
$
--- corresponding to $H_{\text{PFS}}$ --- is available from analysis $k=1$ onward; because no additional PFS events accrue after the interim, the same nominal p-value $p_{\text{PFS}}$ applies at all subsequent analyses.
For $H_{\text{OS}}$, an O'Brien-Fleming-type (OBF) spending function is used across the two analyses, yielding the nominal p-values: $p_{\text{OS},1}$ (OS p-value at interim) and $p_{\text{OS},2}$ (OS p-value at final analysis).

A group sequential graphical procedure is used, and is parameterised by the OBF spending function and an initial graph $(\mathbf{w},\mathbf{G})$.
A two-node graph has a single free parameter governing the alpha-splitting: $w_{\text{PFS}}$ to PFS and $w_{\text{OS}} = 1 - w_{\text{PFS}}$ to OS.
The effective nominal boundaries for OS depend on the current available local significance level (i.e., initial $\alpha^*_{2,k}(w_{\text{OS}} \alpha)$ or  
$\alpha^*_{2,k}(\alpha)$ after alpha is recycled to $H_{\text{OS}}$ upon rejection of $H_{\text{PFS}}$).
As the interim PFS p-value is final, nominal boundaries for PFS depend only on graph recycling and not on a spending function. 

Let $\left(Z_{\text{\text{PFS}}}, Z_{\text{\text{OS}},1}, Z_{\text{\text{OS}},2}\right)$ denote the test statistics for PFS (single statistic available from analysis 1 onward), OS interim, and OS final, respectively. We assume the multivariate normal model
\[
\left(Z_{\text{\text{PFS}}}, Z_{\text{\text{OS}},1}, Z_{\text{\text{OS}},2}\right)^{\top} \sim \mathcal{N}(\boldsymbol \Delta, \boldsymbol \Sigma), \quad
\boldsymbol \Delta=\left(\Delta_{\text{\text{PFS}}}, \Delta_{\text{\text{OS}}} \sqrt{t_{\text{OS},1}}, \Delta_{\text{\text{OS}}}\right)^\intercal,
\]
and use the correlation structure
\[
\operatorname{Corr}\left(Z_{\text{\text{OS}},1}, Z_{\text{\text{OS}},2}\right)=\sqrt{t_{\text{OS},1}}, \quad \operatorname{Corr}\left(Z_{\text{\text{PFS}}}, Z_{\text{\text{OS}},2}\right)=\rho, \quad \operatorname{Corr}\left(Z_{\text{\text{PFS}}}, Z_{\text{\text{OS}},1}\right)=\rho \sqrt{t_{\text{OS},1}},
\]
which corresponds to the group sequential canonical distribution for OS and the parameter $\rho$ for the correlation between the final PFS and OS test statistics. 
We use similar protocol design inputs for the EV-302 trial for our optimisation inputs:
(i) an interim analysis at which PFS and OS are tested once the OS endpoint has reached information fraction $t_{\text{OS},1} = 0.7$, and 
(ii) the noncentrality vector is $\boldsymbol \Delta = (\Delta_\text{\text{PFS}}, \Delta_\text{\text{OS}} \sqrt{t_{\text{OS},1}}, \Delta_\text{\text{OS}})^\intercal = (4.01, 2.87, 3.44)^\intercal$.
These inputs, alongside the value scenarios for Example 5, are presented in Table~\ref{tab:gsd_inputs}.

\begin{table}[!t]
\centering
\caption{Design inputs for Example 5 based on the EV-302 trial.\protect\cite{powles2024, ema_padcev_2024}}
\label{tab:gsd_inputs}
\begin{tabular}{lll}
\hline
\textbf{Component} & \textbf{Value(s)} & \textbf{Notes} \\
\hline
FWER ($\alpha$) & $0.025$ & One-sided \\
Number of analyses ($K$) & $2$ & Interim + final \\
Calendar times $(s_1, s_2)$ & $(37, 47)$ months & From first patient randomised \\
OS information fraction at interim ($t_{\text{OS},1}$) & $0.7$ &  \\
OS spending function & O'Brien--Fleming & \\
PFS ``spending'' & None & Same $p_1$ compared at both looks \\
PFS/OS correlation parameter ($\rho$) & 0.5 & $\operatorname{Corr}(Z_{\text{\text{PFS}}}, Z_{\text{\text{OS}},2})$ \\
NCP under alternative, PFS at interim analysis ($\Delta_{\text{\text{PFS}}}$) & 4.01 & Corresponds to 98\% nominal power at $\alpha=0.025$ \\
NCP under alternative, OS at final analysis ($\Delta_{\text{\text{OS}}}$) & 3.44 & Corresponds to 93\% nominal power at $\alpha=0.025$ \\
\hline
\end{tabular}
\end{table}

We assign value to declared rejections and discount by decision time. Because the gain depends on whether each endpoint is rejected and when a claim can be made, we write it as a function of $\boldsymbol{\tau}= \left(\tau_\text{\text{PFS}}, \tau_\text{\text{OS}}\right):$
\[
\psi(\boldsymbol{\tau})=v_\text{\text{PFS}} d\left(\tau_\text{\text{PFS}}\right)+v_\text{\text{OS}} d\left(\tau_\text{\text{OS}}\right), \quad v_\text{\text{PFS}}+v_\text{\text{OS}}=1,
\]
where $v_\text{\text{OS}} / v_\text{\text{PFS}}$ represents the relative value of OS versus PFS, and where $d(\tau_i)$ is the discount step function mapping decision time (in months from first patient randomised) to a value multiplier:
\[
d(\tau_i) = 
\begin{cases} 
1, & \tau_i = s_1 = 37 \text{ months}, \\
\delta, & \tau_i = s_2 = 47 \text{ months}, \\
0, & \tau_i = \infty.
\end{cases}
\]
(where $i \in \{\text{PFS, OS}\}$) so that a confirmatory claim declared at the final analysis (47 months) is discounted by $\delta$ relative to the same claim declared at the interim analysis (37 months).

Although no spending function is applied to PFS, there is a scenario in which $H_{\text{PFS}}$ is rejected only at the final analysis ($k=2$).  
This would occur if $p_{\text{PFS}} > w_{\text{PFS}} \alpha$ and $p_{\text{OS},1} > \alpha^*_{2,1}(w_{\text{OS}} \alpha)$ at $k=1$, but the observed p-value $p_{\text{OS},2}$ at the final analysis satisfies $p_{\text{OS},2} \leq \alpha^*_{2,2}(w_{\text{OS}} \alpha)$. 
Under these circumstances, alpha recycling from OS to PFS occurs to allow $H_{\text{PFS}}$ to be rejected at the final analysis using the interim nominal PFS p-value, $p_{\text{PFS}}$.
In this case, $\tau_\text{\text{PFS}} = s_2 = 47$ months: the decision time records when the rejection criterion was met, not when the PFS data matured.

In this example, we vary the value ratio $v_\text{\text{OS}} / v_\text{\text{PFS}}$.
By construction, the gain $\psi(\boldsymbol{\tau}) \in[0,1]$ for all possible outcomes.
Under the design inputs of Table~\ref{tab:gsd_inputs}, we evaluate each candidate $w_{\text{PFS}}$ by the conditional expected gain
\[
U\left(w_{\text{PFS}} ; \boldsymbol{\theta}\right)=\mathbb{E}_{\boldsymbol{\theta}}\left[\psi(\boldsymbol{\tau})\right],
\]
by Monte Carlo simulation under $\mathcal{N}(\boldsymbol{\Delta}, \boldsymbol{\Sigma})$.
For each replicate we generate $(Z_{\text{\text{PFS}}}, Z_{\text{\text{OS}},1}, Z_{\text{\text{OS}},2})$, compute $p_{\text{PFS}}, p_{\text{OS},1}, p_{\text{OS},2}$, 
apply the two-hypothesis group sequential graphical procedure to obtain $\boldsymbol{\tau}$,
and average $\psi(\boldsymbol{\tau})$ across replicates.
We reuse the same simulated draws when evaluating $U\left(w_{\text{PFS}}\right)$ across candidate values of $w_{\text{PFS}}$. We evaluate the optimal graph parameter 
\[
w_{\text{PFS}}^{\star} = \argmax_{w_{\text{PFS}} \in[0,1]} U\left(w_{\text{PFS}}\right)
\]
conditional on a given pair $(v_\text{\text{OS}} / v_\text{\text{PFS}}, \delta)$.

\begin{figure}[t]
\centering

\begin{subfigure}[t]{0.49\textwidth}
  \centering
  \includegraphics[width=\linewidth, trim=20 50 20 50, clip]{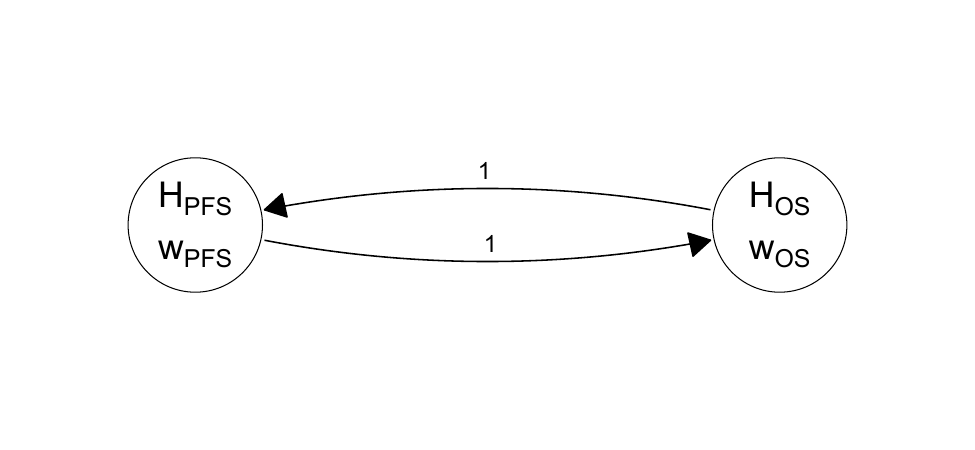}
  \caption{Two-endpoint group sequential graphical procedure for PFS ($H_{\text{PFS}}$) and OS ($H_{\text{OS}}$).}
  \label{fig:gsd_graph}
\end{subfigure}
\hfill
\begin{subfigure}[t]{0.49\textwidth}
  \centering
  \includegraphics[width=\linewidth]{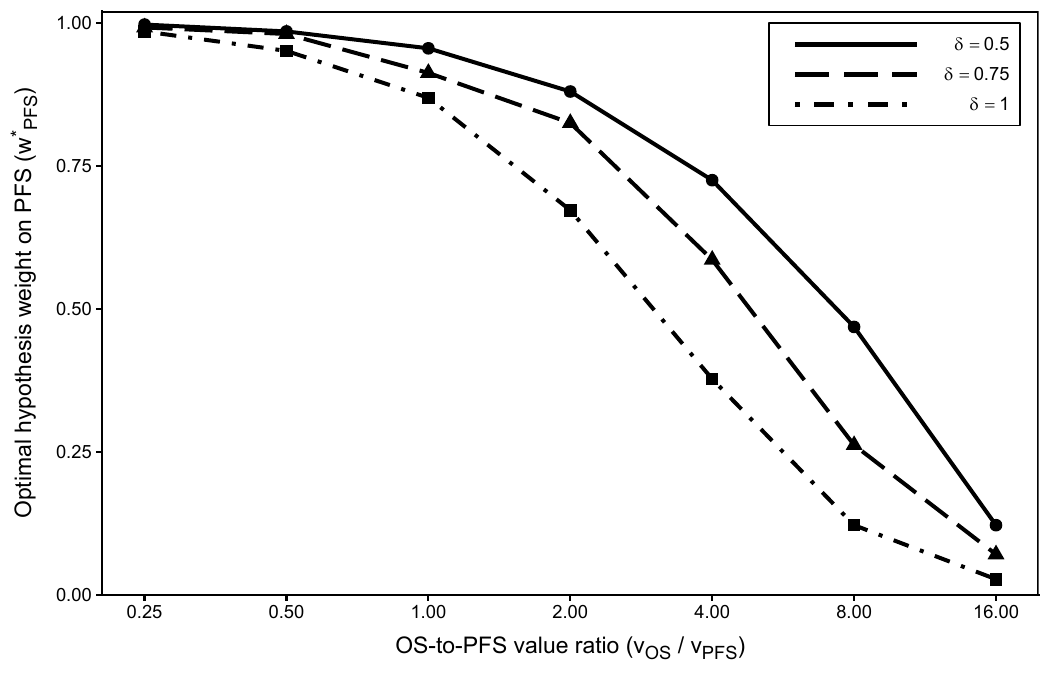}
  \caption{Optimal initial weight on PFS, $w_{\text{PFS}}^*(r,\delta)$, as a function of the OS-to-PFS value ratio $r=v_\text{\text{OS}}/v_\text{\text{PFS}}$ and discount factor $\delta$.}
  \label{fig:gsd_w1}
\end{subfigure}

\caption{Group sequential graphical procedure for Example 5 and resulting optimal initial weight allocation. Panel (a) shows the candidate two-node graph with initial weights $(w_{\text{PFS}},w_{\text{OS}})$ and full recycling between endpoints. Panel (b) shows the optimal initial weight on PFS, $w_{\text{PFS}}^*$, obtained by maximising conditional expected gain under the design inputs in Table~\ref{tab:gsd_inputs}.}
\label{fig:gsd_combined}
\end{figure}

Figure~\ref{fig:gsd_w1} summarises $w_{\text{PFS}}^{\star}(v_\text{\text{OS}} / v_\text{\text{PFS}})$ for discount values $\delta = 1$ (no reduced value if decision time is at final analysis compared to interim), $\delta = 0.75$ and $\delta = 0.5$.
As OS becomes more valuable (larger $v_\text{\text{OS}} / v_\text{\text{PFS}}$), the optimal alpha-splitting shifts toward OS (smaller $w_{\text{PFS}}^{\star}$).
Additionally, when decision-time is discounted more heavily (smaller $\delta$), the optimal allocation shifts toward PFS (larger $w_{\text{PFS}}^{\star}$) for any fixed $v_\text{\text{OS}} / v_\text{\text{PFS}}$.

\section{Discussion}\label{sec:discussion}

While shortcut procedures like the graphical approach are easier to communicate than general closed testing procedures, that ease of communication does not help a trial team decide which graph to use. 
In practice, sponsors default to Holm or fixed-sequence procedures because choosing otherwise can be difficult to justify, not because those defaults are well-matched to the trial's objectives.
The framework developed here replaces that unarticulated default with an objective (gain) function that the trial team has explicitly endorsed. We present it as an internal sponsor tool for selecting a graph at the design stage. 
Since FWER control is a property of the graphical procedure itself, only the chosen graph and its testing algorithm need appear in the protocol.
The gain function itself --- which may encode commercially sensitive judgements about the relative value of different label claims --- need not be disclosed.

The results of Examples 1--5 are consistent with previous findings: when utility-based objectives are used to optimise closed testing procedures, then the optimal procedure depends strongly on how trial success is defined, not only on the assumed effect sizes.\cite{lisovskaja2015} 
However, that work was limited to assuming independent test statistics and identified Bayesian averaging over uncertain parameters as a natural extension without implementing it.
Subsequent frameworks handle correlated test statistics and can accommodate general objective functions, including conditioning secondary power on primary rejection,\cite{zhan2022, xi2024} but little guidance has been published on how to construct the objective itself. Instead, clinical preferences have been incorporated as constraints on graph parameters.\cite{zhang2023}
The hurdle-structured gain function and its elicitation address this directly: stakeholder judgements determine the objective function, and the optimiser identifies the graph that best serves it.
The Bayes gain and time-dependent gain for GSDs then show that the optimal graph depends on how uncertain the assumed treatment effects are (Example~4) and when confirmatory claims become available (Example~5).

In Example~1, the yielded optimised graph was a parallel gatekeeping structure with weighted Holm testing of secondary endpoints within each population before recycling to the other population's primary --- a procedure whose edge weights would have been difficult to elicit directly, but which emerged naturally from the gain function and the assumed joint test statistic distribution.
In Example~2, the submission criterion (dual-endpoint success for at least one dose) is conjunctive within each dose but disjunctive across doses; encoding this directly as the objective, generalising the sample-size formulation of Zhang and Gou,\cite{zhang2023} reduced the required sample size by 10\% relative to a fixed sequence. 
In Example~3, the optimised graph (0.774) outperformed Holm (0.770) by 0.004 in expected gain. The difference may appear small in absolute terms but, for a novel treatment with large total projected revenue, this can represent substantial economic value.\cite{zhan2022, lisovskaja2015}
The computational cost of optimisation is negligible by comparison.
Example~4 shows that the optimal graph depends not only on the assumed treatment effects but on how uncertain those assumptions are. 
Under a fixed alternative the optimal procedure was a fixed sequence; introducing prior uncertainty meant that the optimal graph moved toward greater alpha-recycling among secondary endpoints as the probability of the assumed ordering being correct decreased. 
The independent priors used in that example represent an upper bound on the degree of alpha-recycling; a joint prior on treatment effects would produce less recycling.
We showed in Example~5 that the value of a confirmatory claim depends on when it is declared:
the optimal graph parameters must balance the value of meeting an endpoint against the cost of delayed evidence.
In this PFS/OS oncology example, this trade-off is governed jointly by the relative endpoint value and the discount factor, a dimension that time-invariant power metrics cannot capture.

For practical elicitation, the additive model in Equation~\eqref{eq:additive} assumes that the value of rejecting one hypothesis does not depend on which other incremental hypotheses have been rejected.
In reality, label claims may exhibit diminishing returns; if two endpoints measure related aspects of disease activity, the second claim adds less value once the first is secured.
While the general formulation in Equation~\eqref{eq:hurdle_formula} can accommodate such dependencies, doing so increases the number of parameters to elicit and makes the process correspondingly harder.
The inter-rater reliability of the procedure and the sensitivity of the optimised graph to variability in elicited scores remain areas for formal assessment.

Several limitations apply. The evolutionary algorithm used for examples 1-5 is a metaheuristic and gives no guarantee of finding the global optimum;
the difficulty of finding it grows with the number of hypotheses.
The examples in this paper have $m \leq 6$; for larger families a confirmatory programme may present --- two doses across two populations with several endpoints where there may be 12 or more hypotheses --- the optimisation becomes more demanding and a single run is less likely to locate the global optimum. 
Therefore, we recommend several independent restarts and confirming the best solution on a fresh Monte Carlo sample.
This confirmation also guards against a second risk. 
Simulation sample reuse, which fixes one matrix of p-values throughout the search, is needed for stable convergence, since evaluating each candidate on independently drawn samples injects noise that can lead the optimiser to chase simulation artefacts rather than real improvements;
but reuse lets the returned graph exploit peculiarities of the fixed draws, so the objective value at the returned solution is optimistically biased.
We therefore recommend recomputing the reported gains and local powers on an independent sample.
The fixed simulation draws could also be generated by randomised quasi-Monte Carlo rather than the pseudorandom draws used here, which could reduce the variance of the gain estimates and the simulation budget needed for stable optimisation.\cite{lecuyer2016}
Another issue is that at larger $m$, structurally different graphs may achieve similar expected gain, so that the choice between them is not well determined by the objective alone; the greedy pruning step described in the Supporting Information (Section~S2.1) helps by favouring a parsimonious representative.
However, our implementation evaluates candidate removals one at a time and in a fixed order; it could be strengthened by searching over combinations of simultaneous removals, or by putting a sparsity penalty on the weights and edges directly into the objective function so that the optimiser favours simpler graphs throughout the search rather than only at the end.
More scalable global-search strategies are a direction for future work.

The optimised graph is only as good as the assumptions it is built on. 
A graph optimised using the wrong effect sizes or between-test-statistic correlations may be less powerful than intended. 
Correlations will usually be the hardest of these inputs to obtain at the design stage, since they are rarely reported and are difficult to infer from clinical study reports from related trials.
In Examples 1--5, these inputs were fixed. If there is uncertainty in correlations, this can be accounted for in the optimisation rather than ignored:
see the Supporting Information (Section~S1) for how a prior may be placed on $\boldsymbol{\Sigma}$, so that the graph is chosen to perform well across a plausible range of correlations rather than a fixed value.
The multivariate normal model used throughout is an asymptotic approximation (justified by the central limit theorem for large confirmatory trials) that affects only the gain estimates used to evaluate graphs and not FWER control;
where it is doubtful, any simulable model may be substituted without changing the optimisation.
Misspecifying the gain function yields a graph less well aligned with the sponsor's objectives.\cite{lisovskaja2015}

Several extensions warrant investigation.
Throughout, the graph is optimised with the design held fixed. 
The gain objective also depends on the design parameters, therefore the design and graph could be optimised jointly: for example, treating interim timing and spending parameters as decision variables in a GSD (Section~\ref{sec:gsd_gain}).
The sample-size minimisation of Example 2 could likewise be carried into the sequential setting.
The examples here all use weighted-Bonferroni local tests, but the framework does not depend on this choice.
Simes-based or parametric intersection tests would uniformly improve power and hence gain, since the gain functions used here are monotone in the rejection pattern.
The barrier is computational, since these tests lose consonance, i.e., forfeit the graphical shortcut and require the full closed testing procedure.\cite{bretz2011_biom}
The Bayes gain could also be extended from the independent priors on the standardised treatment effect (used in Example 4) to a joint prior, capturing the co-dependence of effects and correlations and producing more realistic optimal graphs than what was demonstrated.
Finally, the gain function currently depends only on the rejection pattern. Letting it depend on the magnitude of the estimated treatment effects would more closely reflect the value of different label claims, and would accommodate success criteria that pair significance with a bound on the observed effect, as in assurance calculations that condition on a minimum estimated benefit. \cite{lisovskaja2015, hagan2005} 

{\tt multigrain}, an R package implementing the methods in this paper, is available at \href{https://github.com/GSK-Biostatistics/multigrain}{https://github.com/GSK-Biostatistics/multigrain}.
The exact package version used to produce the results, together with the scripts that reproduce the examples of Section~\ref{sec:examples}, are available from the corresponding author on request.


\bmsection*{Conflict of interest}

The authors declare no potential conflict of interests.

\bmsection*{Acknowledgments}

The authors thank Dragoș Moldovan-Grünfeld (GSK) for his major role in developing the {\tt multigrain} R package.

\bmsection*{Data Availability Statement}

No new data were generated in this study; all results derive from simulation.
Continuing development of the {\tt multigrain} R package takes place at \url{https://github.com/GSK-Biostatistics/multigrain}, where later versions may differ from the one used here.
The exact package version used to produce the results, together with the scripts that reproduce every example, figure and table in the paper, are available from the corresponding author on request.

\bibliography{wileyNJD-AMA}

\bmsection*{Supporting information}

The Supporting Information referenced throughout is appended to this preprint
(Sections S1--S3).

\clearpage

\setcounter{section}{0}
\renewcommand{\thesection}{S\arabic{section}}

\begin{center}
{\LARGE\bfseries Supporting Information}
\end{center}
\vskip18pt

\section{Specification of prior distribution for Bayes gain optimisation}

\subsection{Finite mixture priors}

A simple but practical approach to prior specification for computing the Bayes gain is a finite mixture prior that assigns probabilities to a small number of plausible scenarios for $\boldsymbol{\theta}$ jointly.
Let $(\boldsymbol{\Delta}^{(l)}, \boldsymbol{\Sigma}^{(l)})$, $l = 1, \dots, L$, denote $L$ candidate parameter configurations with associated weights $(\omega_1, \dots, \omega_L)$, $\sum_l \omega_l = 1$.
From the results established in Section~\ref{sec:bayes_gain}, the finite mixture induces a predictive density for the p-values:
\[
f^*(\mathbf{p}) = \sum_{l=1}^{L} \omega_l \cdot f\left(\mathbf{p} \mid \boldsymbol{\Delta}^{(l)}, \boldsymbol{\Sigma}^{(l)}\right),
\]
where $f(\mathbf{p} \mid \boldsymbol{\Delta}^{(l)}, \boldsymbol{\Sigma}^{(l)})$ is the p-value density implied by $\mathbf{Z} \sim \mathcal{N}(\boldsymbol{\Delta}^{(l)}, \boldsymbol{\Sigma}^{(l)})$. The Bayes gain is the expected gain under this single predictive distribution:
\[
U_B(\mathbf{w}, \mathbf{G}) = \mathbb{E}_{f^*}\bigl[\psi\left\{R(\mathbf{p}; \mathbf{w}, \mathbf{G})\right\}\bigr].
\]
Finite mixture priors are appealing when a few qualitatively distinct scenarios are considered (e.g., optimistic, expected, and pessimistic effect-size profiles, possibly paired with different correlation structures).

\subsection{Continuous prior on noncentrality parameters}

Sometimes a continuous prior on $\boldsymbol\Delta$ might be preferred; in such cases, we specify a joint multivariate normal prior:
\[
\boldsymbol{\Delta}\sim\mathcal{N}(\boldsymbol{\mu}, \mathbf{V}),
\]
where $\boldsymbol{\mu} = (\mu_1, \dots, \mu_m)^\intercal$ is the prior mean vector and
$\mathbf{V} = \mathbf{D}\mathbf{R}_\Delta\mathbf{D}$, with $\mathbf{D} = \text{diag}(\sigma_{\Delta,1}, \dots, \sigma_{\Delta,m})$
encoding marginal uncertainty in each treatment effect and $\mathbf{R}_\Delta$ the prior correlation matrix for the treatment effects.
The hyperparameters $(\boldsymbol{\mu}, \mathbf V)$
should reflect genuine consensus beliefs about treatment effects at the design phase -- informed by early-phase data, related compounds, and historical success rates -- rather than the fixed point alternative used in conventional power calculations.
This distinction is analogous to that between the ``design prior'' used for assurance and the single alternative used for frequentist power \cite{hagan2005}.
When the possibility of no treatment effect is non-negligible, a mixture prior with mass at zero may be more appropriate than a single normal component.

Under this prior the predictive distribution of the test statistics is 
$\mathbf{Z}\sim\mathcal{N}(\boldsymbol{\mu},\boldsymbol{\Sigma}+\mathbf{V})$. 
Setting $\mathbf{R}_\Delta=\mathbf{I}$ yields independent marginal priors $\Delta_i\sim\mathcal{N}(\mu_i,\sigma_{\Delta,i}^2)$, the special case used in Example~4.
Prior uncertainty then adds to the diagonal of $\boldsymbol{\Sigma}$ but not to the off-diagonal, so predictive correlations are attenuated:
\[
\operatorname{Corr}(Z_i,Z_j)=\frac{\Sigma_{ij}}{\sqrt{(1+\sigma_{\Delta,i}^2)(1+\sigma_{\Delta,j}^2)}}.
\]
Independent priors therefore make the test statistics appear more independent than they are, and the optimised graph can recycle more alpha than the true optimum (see discussion in Example~4).

\subsection{Continuous prior on the correlation matrix}

The correlations between test statistics are often determined by the trial design (e.g., common-control comparisons, nested populations, see Example~1 in Section~\ref{sec:examples}) and can be treated as a fixed input.
When uncertainty in pairwise correlations is important, a prior $\pi_\Sigma(\boldsymbol{\Sigma})$ may be specified on the space of correlation matrices, that is, symmetric positive definite matrices with unit diagonal.
A common choice is the LKJ distribution, with density proportional to $\det(\boldsymbol{\Sigma})^{\eta-1}$: $\eta = 1$ is uniform over the space, while $\eta > 1$ concentrates mass toward the identity~\cite{lewandowski2009}.
Informative beliefs about individual correlations can then be incorporated by placing additional priors (e.g., normal priors centred at the anticipated value) on selected off-diagonal elements of the resulting correlation matrix.
This is straightforward to implement in probabilistic programming frameworks such as Stan.

\section{Optimisation algorithm}
\label{sec:algorithm}

\subsection{Heuristic optimisation procedure}

\subsubsection{Overview}

The optimisation problems in Section~\ref{sec:formulation} of the main text require maximising a Monte Carlo estimate of the expected gain (or Bayes gain) over the graph parameters $(\mathbf{w}, \mathbf{G})$, subject to the regularity constraints of graphical procedures and any additional structural constraints.
The objective function is generally non-convex with many local optima, and optimal solutions frequently lie on the boundary of the feasible set (i.e., some weights or transition probabilities equal zero or one) \cite{xi2024, zhan2022}.
We employ a hybrid evolutionary algorithm (implemented via the R package \texttt{GA}) with Cauchy-distributed mutation and population initialisation operators inspired by the Fast Evolutionary Strategy (FES) of Yao et al.,
combined with intermittent derivative-free Nelder-Mead local search for refinement within promising regions \cite{scrucca2017, yao1997, moscato1989}.

\subsubsection{Simulation sample reuse}\label{sec:ssr}

To evaluate the objective function for any candidate graph, we employ Monte Carlo \textit{simulation sample reuse} \cite{wang2002}.
A single fixed set of $N_{\text{sim}}$ p-value vectors is simulated from the p-value distribution $f(\mathbf{p} \mid \boldsymbol{\theta})$ (or the predictive distribution $f^*$ for Bayes gain) prior to the start of optimisation.
This fixed sample is reused when evaluating every candidate graphical procedure throughout the optimisation run.
Because the same p-value draws are used for all candidates, any observed difference in the estimated objective function is attributable to the change in graph parameters rather than to stochastic simulation noise.
This strategy is analogous to using common random numbers in simulation optimisation and substantially reduces the variance of pairwise comparisons between candidate solutions.
Because the returned solution maximises the estimate over this fixed sample, all expected gains reported in the main text (for optimised graphs and comparator procedures alike) are recomputed on an independent sample not used during optimisation.
The graphical shortcut is applied to each p-value vector, and gain function is calculated as described in Section~\ref{sec:expected_gain}.
To avoid numerical instability in the graphical update step, hypothesis weights below $10^{-4}$ and transition weights below $10^{-5}$ are set to zero before each candidate graph is evaluated.

\subsubsection{Global search}

Candidate procedures are encoded as real-valued vectors in which the last free hypothesis weight, and the last free element of each row of $\mathbf{G}$, are recovered from the corresponding sum constraint.
Structural constraints are imposed by fixing elements of $\mathbf{w}$ or $\mathbf{G}$ in advance; when a graph has no additional constraints beyond the regularity conditions, the encoded vector has dimension $m^2 - m - 1$.
Enforcing equality rather than inequality in the sum constraints excludes no optimal solution, because the gain functions considered here are non-decreasing in the rejection pattern and the rejection pattern is non-decreasing in the hypothesis weights and in the row sums of $\mathbf{G}$, so holding back $\alpha$ can never improve the objective.
Offspring are generated by a multi-parameter mutation operator inspired by the Fast Evolutionary Strategy of Yao et al.~ \cite{yao1997}. Each coordinate is independently selected for mutation with probability $0.1$, and each selected coordinate is replaced by a draw from a Cauchy distribution centred on its current value and truncated to the feasible range.
Refinement is provided separately by the intermittent Nelder-Mead local search and the best solution is protected by elitism, so mutation serves exploration alone; crossover is not used.
Selection follows the GA-package default for real-valued optimisation.

\subsubsection{Intermittent local search}

To accelerate convergence within promising regions identified by the global search, a derivative-free Nelder-Mead local search is applied intermittently.
At each generation, a local search is triggered with probability $p_{\text{optim}} = 0.2$.
When triggered, a single individual $\mathbf{x}_s$ is selected from the current population based on its fitness rank,
and the Nelder-Mead algorithm is initiated from $\mathbf{x}_s$ to find a local optimum $\mathbf{x}_{\text{local}}^*$,
which then replaces $\mathbf{x}_s$ in the population.

\subsubsection{Termination}

The evolutionary algorithm terminates after a pre-specified number of consecutive generations without improvement in the best objective value.
In the examples of the main text, this patience parameter is set to 1000 generations.
A final derivative-free local search (COBYLA, via the \texttt{nloptr} package) is then applied starting from the best solution found, terminating when every parameter changes by less than $5 \times 10^{-8}$ times its absolute value in a single step \cite{nloptr}.

\subsubsection{Post-optimisation graph pruning}

In our experience, the above optimisation procedure frequently returned near-zero hypothesis weights or transition probabilities that carry negligible influence on expected gain at the cost of a more complicated graph.
A pruning step is therefore applied to the final solution:
each weight $w_i$ and each edge $g_{ij}$ is tentatively set to zero in turn (with the remaining weights or outgoing edges renormalised), and the removal is accepted if the estimated gain does not decrease.
Removals are evaluated one parameter at a time rather than over all subsets, and accepted removals are carried forward so that later candidates are evaluated on the already-pruned graph.
The result is therefore order-dependent and not guaranteed to be the sparsest graph achieving the optimal gain.
Because removals are evaluated greedily, this returns a simpler graph with the same estimated gain or higher.
Pruning decisions are made on the optimisation sample, and the pruned graph is re-evaluated on the independent sample described in Section~\ref{sec:ssr}.

\subsection{Adaptation for sample-size minimisation}

When minimising sample size (Section~\ref{sec:formulation} of the main text), the p-value distribution changes with the per-arm sample size $n$, preventing direct reuse of a fixed p-value matrix.
To retain the variance-reduction benefits of sample reuse, we instead fix the underlying simulation noise: a matrix $\mathbf{Z}_{\text{null}}$ is generated once from $\mathcal{N}(\mathbf{0}, \boldsymbol{\Sigma})$, and test statistics for any candidate $n$ are constructed as $\mathbf{Z} = \mathbf{Z}_{\text{null}} + \boldsymbol{\Delta}(n)$, yielding p-values $p_i = 1 - \Phi(Z_i)$.

The core algorithm is wrapped in a two-phase procedure to handle the integer constraint on $n$ and the non-linear constraint on expected gain. We also define a signed fitness function that penalises infeasible solutions (i.e., when the estimated conditional expected gain $\hat{U}$ falls below the target expected gain $\gamma$):
\[
f(\mathbf{w}, \mathbf{G}, n) =
\begin{cases}
U - \gamma & \text{if } U \geq \gamma \quad \text{(feasible)} \\
-(1 + \lvert U - \gamma \rvert) & \text{if } U < \gamma \quad \text{(infeasible)}
\end{cases}
\]
which ensures feasible solutions strictly dominate infeasible ones.

\begin{itemize}
    \item \textbf{Phase 1 (Bisection).} A binary search over $n$ is conducted using a small fixed set of seed graphs (e.g., fixed sequence, Holm-Bonferroni) to rapidly identify an initial feasible sample size $n_{\text{init}}$.
    \item \textbf{Phase 2 (Step-down).} Starting at $n_{\text{init}}$, the evolutionary algorithm runs successive blocks of $\kappa$ generations at fixed $n$.
    After each block, if the current best solution remains feasible at $n - 1$, $n$ is decremented (repeatedly, if still feasible at further reductions) and the population is re-evaluated at the new sample size.
    This focuses computational effort on finding the best-performing graph at the boundary of feasibility.
\end{itemize}

\subsection{Extension to group sequential designs}

When information fractions and spending functions are treated as fixed inputs, optimisation of graphs within group sequential designs (GSDs) differs from the fixed-sample case only in how the objective function is evaluated.
The decision variables remain the initial graph parameters $(\mathbf{w}, \mathbf{G})$, so the dimension of the search space and the encoding of candidate solutions are unchanged.

For $m$ hypotheses and $K$ analyses, let $t_{i,k}$ denote the information fraction for $H_i$ at analysis $k$ and $\rho_{ij}$ the correlation between the final test statistics for $H_i$ and $H_j$. Under the canonical joint distribution,
\[
\mathbb{E}[Z_{i,k}] = \Delta_i\sqrt{t_{i,k}}, \qquad \operatorname{Corr}(Z_{i,k}, Z_{j,l}) = \rho_{ij}\sqrt{\frac{\min(t_{i,k}, t_{j,l})}{\max(t_{i,k}, t_{j,l})}},
\]
with $\rho_{ii} = 1$, which reduces to $\sqrt{t_{i,k}/t_{i,l}}$ within a hypothesis. 
To implement simulation sample reuse, a fixed array $\{Z_{i,k}^{(b)}\}$ of dimension $N_{\text{sim}} \times m \times K$ is simulated once from this distribution and reused for every candidate graph, with nominal one-sided p-values $p_{i,k}^{(b)} = 1 - \Phi(Z_{i,k}^{(b)})$ at each analysis.
During each evaluation step, the group sequential graphical procedure is applied to these fixed sequences using the candidate graph and the pre-specified spending functions.
This determines the rejection-time vector $\boldsymbol{\tau}$ for each replicate, from which the time-dependent gain $\psi(\mathbf{r}, \boldsymbol{\tau})$ is computed.

Example~5 has $m = 2$ and $K = 2$, and the transition matrix is fixed at full recycling in both directions, so the only decision variable is the initial weight $w_{\text{PFS}} \in [0,1]$.
The heuristic search described above is unnecessary for a one-dimensional problem and was not used: the objective was instead evaluated on a dense grid over $[0,1]$, so the reported optimum is global up to the grid resolution.

\section{Additional example gain functions in the group sequential design case}

In the main body of the paper, Example~5 included a description of a simple potential gain function for a trial evaluating PFS and OS group sequentially.
Here, we expand and include additional examples of useful constructions of the gain function in the group sequential case.

First, we consider how a gain function may be specified for co-primary or dual endpoints analysed at multiple time points.
For simplicity, consider a scenario in which the two endpoints, denoted $1$ and $2$, are tested at two time points, $s_1$ and $s_2$.
In either instance, a key consideration may then be the requirement, or added value, of achieving both endpoints at the same analysis.
Beginning with dual primary endpoints, a general formulation of the gain function may be:

\[
\psi(\boldsymbol{\tau}) = \nu_1d_1(\tau_1) + \nu_2d_2(\tau_2) + \nu_{12}d_{12}(\tau_1, \tau_2).
\]

Here, $\nu_i$ and $d_i(\tau_i)$ are the value and discount function for endpoint $i = 1, 2$.
Typically, it would suffice to have:

\[
d_i(\tau_i) = \begin{cases}
                 1,        & \tau_i = s_1,   \\
                 \delta_i, & \tau_i = s_2,   \\
                 0,        & \tau_i = \infty.
               \end{cases}
\]

Thus, $\nu_1d_1(\tau_1) + \nu_2d_2(\tau_2)$ allows value to be realised in the case either endpoint achieves significance.
Often, it may be the case that $\nu_1 = \nu_2$ and $d_1(\tau_1) = d_2(\tau_2)$ is a reasonable restriction.

The additional component $\nu_{12}d_{12}(\tau_1, \tau_2)$ represents an added value in the case that both endpoints are significant at the same analysis (which may, e.g., allow a single submission to realise both endpoints in the label).
Again, some special case of the following construction should reflect practical reality:

\[
d_{12}(\tau_1, \tau_2) = \begin{cases}
                              1,           & \tau_1 = \tau_2 = s_1, \\
                              \delta_{12}, & \tau_1 = \tau_2 = s_2, \\
                              0,           & \text{otherwise}.
                            \end{cases}
\]

We may also often expect that $\nu_{12}$ is small relative to $\nu_{1}$ and $\nu_2$.

By contrast, in the co-primary case, we need to account for the fact that no value is realised unless significance is (eventually) achieved for both endpoints.
If the endpoints need to achieve significance at the same time, we may simply use:

\[
\psi(\boldsymbol{\tau}) = \nu_{12}d_{12}(\tau_1, \tau_2),
\]

with $d_{12}(\tau_1, \tau_2)$ taking the form shown above.
If instead value can be realised through significance at different time points, we may extend the above to:

\[
\psi(\boldsymbol{\tau}) = \nu_{12}d_{12}(\tau_1, \tau_2) + \tilde{\nu}_{12}\mathbb{I}\{\tau_1 = s_1\}\mathbb{I}\{\tau_2 = s_2\} + \bar{\nu}_{21}\mathbb{I}\{\tau_1 = s_2\}\mathbb{I}\{\tau_2 = s_1\}.
\]

Here, $\tilde{\nu}_{12}$ is the value from achieving endpoint~1 at analysis~1 and endpoint~2 at analysis~2, while $\bar{\nu}_{12}$ is the value from achieving endpoint~1 at analysis~2 and endpoint~2 at analysis~1.
Often, $\tilde{\nu}_{12} = \bar{\nu}_{12}$ may be a reasonable constraint.

To conclude, we consider a more complex gain function for the scenario in which two endpoints are tested with one expected to allow realisation of accelerated approval and the other full approval.
For ease of understanding, we will refer to these as PFS and OS, reflecting a common role for these two endpoints.
As in Example~5, we assume that $t_{\text{PFS},1} = t_{\text{PFS},2} = t_{\text{OS},2} = 1$ and $t_{\text{OS},1} < 1$.
Because the PFS p-value is unchanged between analyses, PFS significance at the final analysis requires alpha-recycling to have occurred from OS at that analysis; the combination $\tau_{\text{OS}} = s_1$, $\tau_{\text{PFS}} = s_2$ therefore cannot arise, and no corresponding term is needed below.
Then, we may like our gain function to reflect the following considerations:

\begin{itemize}
    \item If OS is significant at the interim analysis, then its value should be higher than if it is significant at the second analysis. PFS should have value in this scenario, but it may be low because OS has achieved significance.
    \item If OS is significant at the final analysis, its value is lower than in the scenario above. PFS significance at the interim analysis is now worth far more; PFS significance at the final analysis may be worth little as again an OS result has been achieved at the same time.
    \item If OS is not significant, then PFS significance at the interim analysis may carry some small value (e.g., if there is a long time period between the interim and final analyses, during which accelerated approval would be in place, before potentially being withdrawn following the final OS analysis).
\end{itemize}

This could be realised through a gain function of the following form:

\begin{equation*}
    \begin{aligned}
        \psi(\boldsymbol{\tau}) &= \mathbb{I}\{\tau_\text{OS} = s_1\}(\nu_\text{OS,1} + \nu_{\text{PFS,1}}\mathbb{I}\{\tau_\text{PFS} = s_1\}) \\ & \qquad + \mathbb{I}\{\tau_\text{OS} = s_2\}(\nu_\text{OS,2} + \nu_{\text{PFS,1}}'\mathbb{I}\{\tau_\text{PFS} = s_1\} + \nu_{\text{PFS,2}}\mathbb{I}\{\tau_\text{PFS} = s_2\}) \\ & \qquad + \mathbb{I}\{\tau_\text{OS} = \infty\}\mathbb{I}\{\tau_\text{PFS} = s_1\}\nu_{\text{PFS},1}'').
    \end{aligned}
\end{equation*}

The three terms here reflect value in relation to OS significance at the interim analysis, final analysis, and not achieving OS significance, as described in the three bullet points above.
Often, it will be the case that $\nu_{\text{PFS,1}}' > \nu_{\text{PFS,1}} > \nu_{\text{PFS},1}''$.
Note that this can be reparameterised as:

\[
\psi(\boldsymbol{\tau}) = \nu_\text{OS}d_\text{OS}(\tau_\text{OS})\{ 1 + \nu_\text{PFS}d_\text{PFS}(\tau_\text{PFS}, \tau_\text{OS}) \},
\]

visually reflecting the (discounted) value of when we achieve significance on OS (constructing $d_\text{OS}(\cdot)$ such that $d_\text{OS}(\infty) > 0$), and the added value of PFS significance.

\end{document}